\documentclass[paper=A4,11pt]{scrartcl}
\pdfoutput=1
\usepackage{graphicx}
\usepackage{tabularx}
\usepackage{array}
\usepackage{amsmath}
\usepackage{amssymb}
\usepackage{color}
\usepackage{longtable}
\usepackage{verbatim}
\usepackage{cite}
\usepackage{bbm}
\usepackage[normalem]{ulem}
\usepackage[bookmarks=false]{hyperref}

\addtokomafont{disposition}{\rmfamily\boldmath}

\newcommand{\beq}{\begin{equation}}
\newcommand{\eeq}{\end{equation}}
\newcommand{\ba}{\begin{array}}
\newcommand{\ea}{\end{array}} 
\newcommand{\beqa}{\begin{eqnarray}}
\newcommand{\eeqa}{\end{eqnarray}}

\def \eff{\rm eff}

\newcommand{\bdmm}{B_d\to\mu^+\mu^-}
\newcommand{\bsmm}{B_s\to\mu^+\mu^-}
\newcommand{\bstt}{B_s\to\tau^+\tau^-}

\newcommand{\brbsmm}{{\rm BR}(B_s\to\mu^+\mu^-)}

\begin{document}

\thispagestyle{empty}
\setcounter{page}{0}

\begin{flushright}
\begin{tabular}{l}
CERN-PH-TH/2011-275\\
FERMILAB-PUB-11-597-T
\end{tabular}
\end{flushright}
\vskip1.5cm

\begin{center}
{\LARGE \bf \boldmath Model-independent constraints\\[0.1cm] on new physics in $b\to s$ transitions}\\[0.8 cm]
{\large
Wolfgang~Altmannshofer$^{a}$, Paride~Paradisi$^{b}$ and David M. Straub$^{c}$} \\[0.5 cm]
\small
$^a$ {\em Fermi National Accelerator Laboratory, P.O. Box 500, Batavia, IL 60510, USA}\\[0.1cm]
$^b${\em CERN, Theory Division, CH-1211, Geneva 23, Switzerland} \\[0.1cm]
$^c${\em Scuola Normale Superiore and INFN, Piazza dei Cavalieri 7, 56126 Pisa, Italy}
\end{center}

\bigskip
\begin{abstract}
We provide a comprehensive model-independent analysis of rare decays involving the
$b\to s$ transition to put constraints on dimension-six $\Delta F=1$ effective operators.
The constraints are derived from all the available up-to-date experimental data from
the B-factories, CDF and LHCb. The implications and future prospects for observables
in $b \to s \ell^+ \ell^-$ and $b\to s\nu\bar\nu$ transitions in view of improved 
measurements are also investigated.
The present work updates and generalises previous studies providing, at the same time,
a useful tool to test the flavour structure of any theory beyond the SM.
\end{abstract}
\vfill

\newpage
\pagenumbering{arabic}

\section{Introduction}

The CKM description of flavour and CP violation in the Standard Model (SM) has been extremely successful in describing the data from high-precision flavour experiments,
among them the $B$ factories and the Tevatron, and is also in agreement with recent measurements of rare $B$ decays at LHC.
This is a remarkable fact, given that the origin of flavour, i.e. the origin of the replication of fermion species and of the hierarchies in their masses and mixing is still a complete mystery. If one supposes new physics (NP) to be at work not too far above the electroweak scale, as is implied by the gauge hierarchy problem, this fact is even more surprising, since NP coupling to the SM generally leads
to modifications of flavour violation, in particular in flavour-changing neutral current (FCNC) processes, which are suppressed by the GIM mechanism and arise only at the loop level in the SM. Even under the most restrictive of assumptions that can be made on the flavour sector of a NP theory, namely assuming all flavour violation to be governed by the SM Yukawa couplings (the hypothesis of Minimal Flavour Violation \cite{Buras:2000dm,D'Ambrosio:2002ex}, MFV), many models predict significant deviations in FCNC observables.
Moreover, many of the flavour-violating couplings are rather poorly constrained as yet and still allow for sizable NP contributions.
In $\Delta B=1$ transitions, this is particularly true for NP contributions with a different CP phase or chirality with respect to the SM contribution. In that case, large deviations in observables {\em not} measured yet are still easily possible. Quantifying this statement is one of the main goals of this paper.

Experimentally, the prospects to improve constraints on flavour violating couplings are excellent: Today, the LHCb experiment has a high sensitivity to exclusive hadronic, semi-leptonic and leptonic $B$ and $B_s$ decays \cite{:2009ny}.\footnote{In the case of leptonic decays, also ATLAS and CMS are competitive.}
In the mid-term future, two next-generation $B$ factories will allow also the measurement of inclusive rare decays and decays with neutrinos in the final state \cite{Aushev:2010bq,O'Leary:2010af}.

Since FCNC processes test NP indirectly, through quantum corrections induced by heavy particles, the impact of NP on low-energy observables like branching ratios or asymmetries can be summarised by their modification of Wilson coefficients of local, non-renormalizable operators. These short-distance coefficients can be constrained on a completely model-independent basis by measuring FCNC observables and these constraints can in turn be used to constrain individual NP models.

In this paper, we concentrate on $\Delta B=\Delta S=1$ processes, i.e. rare decays with a $b\to s$ transition.
We use up-to-date experimental constraints -- in particular, we include the recent measurement at LHCb of angular observables in $B\to K^*\mu^+\mu^-$ \cite{LHCb-CONF-2011-038}, which is the most precise to date -- to put model-independent constraints on the Wilson coefficients.\footnote{Instead, we do not consider here observables related to the decay $B \to K \ell^+ \ell^-$, since their current experimental resolutions are rather poor. Including them would affect our results only in a negligible way. However, once LHCb data on $B \to K \ell^+ \ell^-$ observables will become available, it will be important to include them given their high NP sensitivity \cite{Bobeth:2007dw,Bobeth:2011nj}.}
Similar constraints have been considered in the literature before, e.g. in the context of MFV \cite{Bobeth:2005ck,Haisch:2007ia,Hurth:2008jc}, magnetic penguin operators \cite{DescotesGenon:2011yn}, SM operators \cite{Gambino:2004mv,Bobeth:2010wg,Bobeth:2011gi} or generic NP \cite{Hiller:2003js,Bobeth:2008ij,Bobeth:2011st}.
We generalise these studies by considering SM operators, their chirality-flipped counterparts
and generic CP violation and considering also the case where they are all simultaneously present instead of considering only a pair at a time.
After imposing the above constraints, we investigate the room left for NP in observables which have not been measured yet, like the branching ratios of $B_s\to\mu^+\mu^-$ and $B_s\to\tau^+\tau^-$, CP asymmetries in $B\to K^*\mu^+\mu^-$ and observables in $b \to s \nu\bar\nu$ decays.

The paper is organised as follows. In section~\ref{sec:obs}, we specify the effective Hamiltonian for $b\to s$ transitions and discuss all the observables relevant for our study. In section~\ref{sec:modelind}, we present our numerical results for the model-independent constraints on the Wilson coefficients. In section~\ref{sec:zpeng}, we consider the constraints in the more restrictive case of semi-leptonic operators generated dominantly by modified $Z$ couplings, which is the case in many models beyond the SM. This allows us to correlate $b\to s\ell^+\ell^-$ processes to $b\to s\nu\bar\nu$ processes. Our main findings are summarised in section~\ref{sec:concl}.

\section{Observables in $b\to s\gamma$ and $b\to s\ell^+\ell^-$ decays}
\label{sec:obs}

The effective Hamiltonian relevant for $b\to s\gamma$ and  $b\to s\ell^+\ell^-$
transitions is given by~\cite{Bobeth:1999mk,Bobeth:2001jm}
\begin{equation}
\label{eq:Heff}
{\cal H}_{\eff} = - \frac{4\,G_F}{\sqrt{2}} V_{tb}V_{ts}^* \frac{e^2}{16\pi^2}
\sum_i
(C_i O_i + C'_i O'_i) + \text{h.c.}~.
\end{equation}
The operators $O_i$ that are most sensitive to NP effects are
\begin{align}
\label{eq:O7}
O_7 &= \frac{m_b}{e}
(\bar{s} \sigma_{\mu \nu} P_R b) F^{\mu \nu},
&
O_8 &= \frac{g m_b}{e^2}
(\bar{s} \sigma_{\mu \nu} T^a P_R b) G^{\mu \nu \, a},
\nonumber\\
O_9 &= 
(\bar{s} \gamma_{\mu} P_L b)(\bar{\ell} \gamma^\mu \ell)\,,
&
O_{10} &=
(\bar{s} \gamma_{\mu} P_L b)( \bar{\ell} \gamma^\mu \gamma_5 \ell)\,,
\nonumber\\
O_S &= 
m_b (\bar{s} P_R b)(  \bar{\ell} \ell)\,,
&
O_P &=
m_b (\bar{s} P_R b)(  \bar{\ell} \gamma_5 \ell)\,,
\end{align}
where $m_b$ denotes the running $b$ quark mass in the $\overline{\rm MS}$
scheme and $P_{L,R}=(1\mp\gamma_5)/2$.
The corresponding operators $O'_i$ are obtained from the operators $O_i$ via
the replacement $P_{L} \leftrightarrow P_{R}$. Numerical values of the Wilson
coefficients and the relation of the coefficients at the matching scale to
the effective low-energy coefficients are discussed in appendix~\ref{sec:ceff}.

Note that we assume $C^{(\prime)}_{9}$ and $C^{(\prime)}_{10}$
to be independent of the lepton flavour, but $C^{(\prime)}_{S}$ and $C^{(\prime)}_{P}$ to be
proportional to the lepton Yukawa couplings (this assumption will become relevant when we discuss the relation between $B_s \to \mu^+\mu^-$ and $B_s \to \tau^+\tau^-$).
Moreover, we also assume lepton flavour conservation, which is an excellent approximation for our purposes, given the stringent experimental bounds on lepton flavour violating processes.

We will now discuss the observables in processes sensitive to this effective
Hamiltonian, which can be used to constrain new physics.

\subsection{$B_s\to\ell^+\ell^-$}\label{sec:bsmm}

In the SM, the $\bsmm$ decay is strongly helicity suppressed and among the rarest FCNC decays.
Using directly the value of the $B_s$ meson decay constant from the lattice,
$f_{B_s} = (250\pm12)$~MeV~\cite{Laiho:2009eu} we get the following SM prediction\footnote{Using the remarkably precise value for $f_{B_s}$ obtained very recently in~\cite{McNeile:2011ng}: $f_{B_s} = (225\pm4)$~MeV, we obtain $\text{BR}(B_s\to\mu^+\mu^-)_\text{SM}=(3.0\pm0.2)\times10^{-9}$~.
This result is as precise as the value that is obtained by assuming $\Delta M_s$ free of NP and using its measurement to reduce the theory uncertainty in $\bsmm$ arising from the $B_s$ meson decay constant~\cite{Buras:2003td}. However, to be conservative, we use~(\ref{eq:bsmm_SM2}) in our analysis.}
\begin{equation} \label{eq:bsmm_SM2}
\text{BR}(B_s\to\mu^+\mu^-)_\text{SM}=(3.7\pm0.4)\times10^{-9} \,.
\end{equation}
The LHCb and CMS collaborations have set a combined upper bound on
the branching ratio of~\cite{LHCb-CONF-2011-037,Chatrchyan:2011kr,CMS-PAS-BPH-11-019}
\beq \label{eq:bsmm_bound}
\brbsmm_\text{LHC}<1.1\times 10^{-8}
\eeq
at 95\% confidence level.\footnote{We mention that the CDF collaboration found an excess of candidates for $\bsmm$
decays, which have been used to determine~\cite{Aaltonen:2011fi}
$\brbsmm_\text{CDF}=\left(1.8^{+1.1}_{-0.9}\right)\times 10^{-8}$~.}

In a generic NP model, the branching ratio of ${\rm BR}(B_s\to\mu^+\mu^-)$ is given by
\begin{equation}
\frac{{\rm BR}(B_s\to\mu^+\mu^-)}{{\rm BR}(B_s\to\mu^+\mu^-)_{\rm SM}} =
|S|^2 \left( 1 - \frac{4 m_\mu^2}{m_{B_s}^2} \right) + |P|^2,
\label{eq:BRbsmm}
\end{equation}
where
\begin{equation}
S = \frac{m_{B_s}^2}{2 m_\mu} \frac{(C_S - C_S')}{|C^{\rm SM}_{10}|} ,
\qquad
P = \frac{m_{B_s}^2}{2 m_\mu} \frac{(C_P - C_P')}{C^{\rm SM}_{10}} +
\frac{(C_{10} - C_{10}')}{C^{\rm SM}_{10}}.
\label{eq:SP}
\end{equation}
The important feature of $\bsmm$ as a probe of NP is that it is among the very few
$b\to s$ decays that are strongly sensitive to scalar and pseudoscalar operators.
In models where such operators are sizable, the branching ratio can easily saturate
the experimental limit. In section~\ref{sec:modelind}, we will use measurements of
other $b\to s$ processes to constrain the Wilson coefficients $C^{(\prime)}_{10}$,
which then allows us to predict the maximum allowed size of BR($\bsmm$)
{\em in the absence} of (pseudo)scalar currents.

The process $\bstt$,
is governed by Wilson coefficients analogous to $\bsmm$ and its branching ratio is given
by eqs.~(\ref{eq:BRbsmm}) and~(\ref{eq:SP}) with the appropriate replacement $\mu\to\tau$.

\subsection{$b\to s\gamma$}

The experimental data on the branching ratio of the inclusive $B \to X_s \gamma$ decay~\cite{Asner:2010qj}
\begin{equation} \label{eq:bsgamma_exp}
{\rm BR}(B \to X_s \gamma)_{\rm exp} = (3.55 \pm 0.26) \times 10^{-4}~,
\end{equation}
and the corresponding NNLO SM prediction~\cite{Misiak:2006zs,Misiak:2006ab}
\begin{equation} \label{eq:bsgamma_SM}
{\rm BR}(B \to X_s \gamma)_{\rm SM} = (3.15 \pm 0.23) \times 10^{-4}~,
\end{equation}
show good agreement. As the BR$(B \to X_s \gamma)$ is highly sensitive to NP contributions
to the Wilson coefficients $C_7^{(\prime)}$, this agreement leads to severe constraints on
the flavour sectors of many NP models.
In our numerical analysis, we use the expression for the branching ratio reported in~\cite{Lunghi:2006hc}, rescaled to the SM prediction of~\cite{Misiak:2006zs},
and assume the uncertainty in the SM prediction as relative error on the theory prediction.

Another interesting observable that probes the $b \to s \gamma$ transition is the time-dependent
CP asymmetry in the exclusive $B_d \to K^* (\to K^0_S \pi^0) \gamma$ decay \cite{Atwood:1997zr,Grinstein:2004uu,Ball:2006cva,Ball:2006eu}
\begin{equation} \label{eq:def_SKstargamma}
\frac{\Gamma(\bar B^0(t) \to \bar K^{*0}\gamma) - \Gamma(B^0(t) \to K^{*0}\gamma)}{\Gamma(\bar B^0(t) \to \bar K^{*0}\gamma) + \Gamma(B^0(t) \to K^{*0}\gamma)} = S_{K^*\gamma} \sin(\Delta M_d t) - C_{K^*\gamma} \cos(\Delta M_d t)~.
\end{equation}
The coefficient $S_{K^*\gamma}$ is highly sensitive to right handed currents as at leading order it vanishes
for $C_7^\prime \to 0$. As a consequence, SM contributions to $S_{K^*\gamma}$ are suppressed by $m_s/m_b$ or $\Lambda_{\rm QCD} / m_b$~\cite{Grinstein:2004uu}, resulting in a very small SM prediction~\cite{Ball:2006eu}
\begin{equation}
S_{K^*\gamma}^{\rm SM} = (-2.3 \pm 1.6)\%~.
\label{eq:SKgSM}
\end{equation}
Experimental evidence for a large $S_{K^*\gamma}$ would be a clear indication of NP effects
through right handed currents.
On the experimental side one has presently \cite{Ushiroda:2006fi,Aubert:2008gy,Asner:2010qj}
\begin{equation}
S_{K^*\gamma}^{\rm exp} = -0.16 \pm 0.22
\end{equation}
and the prospects to improve this measurement significantly at next generation B factories are excellent~\cite{Meadows:2011bk}.
In our numerical analysis we use the LO expression for $S_{K^*\gamma}$~\cite{Ball:2006cva}
\begin{equation} \label{eq:SKstargamma_NP}
S_{K^*\gamma} \simeq \frac{2}{|C_7|^2 + |C_7^\prime|^2} {\rm Im}\left( e^{-i\phi_d} C_7 C_7^\prime\right)~,
\end{equation}
that leads to accurate predictions in presence of NP.
In the above expression for $S_{K^* \gamma}$, $\sin(\phi_d) = S_{\psi K_S}$ is the phase of the $B_d$ mixing amplitude and the Wilson coefficients are evaluated at the scale $\mu = m_b$. Using directly the experimental value
for $S_{\psi K_S} = 0.67 \pm 0.02$~\cite{Asner:2010qj}, we automatically capture possible
NP effects in $B_d$ mixing. In our NP analysis, we assume the theory uncertainty to be equal
to the SM uncertainty in (\ref{eq:SKgSM}).

Another observable that is in principle sensitive to CP violating effects in the $b \to s \gamma$ transition
is $A_{\rm CP}(b\to s\gamma)$, the direct CP asymmetry in the $B \to X_s \gamma$ decay~\cite{Soares:1991te,Kagan:1998bh}.
In contrast to the observables discussed so far, it is also highly sensitive to NP contributions
to the Wilson coefficients $C_8^{(\prime)}$. However, as shown in~\cite{Benzke:2010tq}, the SM prediction for $A_{\rm CP}(b\to s\gamma)$ is dominated by long-distance contributions and large hadronic uncertainties make it difficult to predict this observable reliably in the context of
NP scenarios. We therefore do not consider $A_{\rm CP}(b\to s\gamma)$ in our analysis.
We also do not consider the isospin asymmetry in $B \to K^* \gamma$, as large hadronic uncertainties strongly limit the constraining power of this observable.

\subsection{$B\to X_s\ell^+\ell^-$}

We consider the inclusive $B\to X_s\ell^+\ell^-$ decay in two different regions of the dilepton invariant mass. The low $q^2$ region with 1~GeV$^2 < q^2 < 6$~GeV$^2$ and the high $q^2$ region with $q^2 > 14.4$~GeV$^2$. Averaging the available results from BaBar~\cite{Aubert:2004it} and Belle~\cite{Iwasaki:2005sy} one finds the following averages for the branching ratios in the two regions
\begin{equation}
{\rm BR}(B \to X_s \ell^+ \ell^-)_{[1,6]}^{\rm exp} = (1.63 \pm  0.50)\, 10^{-6}~,~~ {\rm BR}(B \to X_s \ell^+ \ell^-)_{>14.4}^{\rm exp} = (4.3 \pm 1.2)\, 10^{-7} ~.
\end{equation}
These results should be compared to the SM predictions
\cite{Bobeth:1999mk,Asatrian:2001de,Asatryan:2001zw,Bobeth:2003at,Ghinculov:2003qd,Huber:2005ig,Ligeti:2007sn,Huber:2007vv,Greub:2008cy}
\begin{equation}
{\rm BR}(B \to X_s \ell^+ \ell^-)_{[1,6]}^{\rm SM} = (1.59 \pm  0.11) \, 10^{-6}~,~~ {\rm BR}(B \to X_s \ell^+ \ell^-)_{>14.4}^{\rm SM} = (2.3 \pm 0.7) \, 10^{-7} ~.
\end{equation}
While the low $q^2$ values are in perfect agreement, the SM prediction in the high $q^2$ region is on the low side of the experimental result.
We remark that the theory prediction in the low $q^2$ region quoted above does not include effects from the experimental cut on the hadronic final state. Such effects can be as large as 10\%~\cite{Lee:2005pwa,Bell:2010mg}. Enlarging the theory error correspondingly however would effect our results only in a minor way, given the huge experimental uncertainty in BR$(B \to X_s \ell^+ \ell^-)$.

The branching ratio in the high $q^2$ region is mainly sensitive to NP contributions to the Wilson coefficients $C_9^{(\prime)}$ and $C_{10}^{(\prime)}$, while the branching ratio in the low $q^2$ region also depends strongly on $C_7^{(\prime)}$. In our numerical analysis, we use the expressions given in~\cite{Hurth:2008jc}, adjusting them to take into account also the primed Wilson coefficients, and treat the uncertainties in the SM predictions as relative errors on the theory predictions. 

In principle, another interesting observable to constrain NP would be the for\-ward-back\-ward asymmetry in $B\to X_s\ell^+\ell^-$. However, since it has not been measured yet, we do not include it in our analysis. A fully inclusive measurement is probably not feasible at LHCb and it will only be possible at next generation $B$ factories \cite{Browder:2008em,O'Leary:2010af,Meadows:2011bk}.

\subsection{$B\to K^*\mu^+\mu^-$}
\label{sec:bksmumu}

\begin{table}
\begin{center}
\renewcommand{\arraystretch}{1.2}
\begin{tabular}{c|ccccc|l}
Obs.  & \cite{Altmannshofer:2008dz} & \cite{Egede:2008uy} & \cite{Bobeth:2008ij} & \cite{:2008ju,:2009zv,Aaltonen:2011cn} & \cite{Aaltonen:2011ja} & most sensitive to\\
\hline
$F_L$ & $-S_2^c$ & $F_L$ && $F_L$ & $F_L$ & $C_{7,9,10}^{(\prime)}$ \\
$A_\text{FB}$ & $\frac{3}{4}S_6^s$ & $A_\text{FB}$ & $A_\text{FB}$ & $-A_\text{FB}$ & $-A_\text{FB}$ & $C_7, C_9$ \\
$S_5$ & $S_5$ &&  &  && $C_7,C_7',C_9,C_{10}'$\\
$S_3$ & $S_3$ & $\frac{1}{2}(1-F_L)A_T^{(2)}$ &&& $\frac{1}{2}(1-F_L)A_T^{(2)}$ & $C_{7,9,10}^{\prime}$\\
$A_9$ & $A_9$ && $\frac{2}{3}A_9$ &  & $A_{im}$ & $C_{7,9,10}^{\prime}$ \\
$A_7$ & $A_7$ && $-\frac{2}{3}A_7^D$ &  && $C_{7,10}^{(\prime)}$
\end{tabular}
\renewcommand{\arraystretch}{1.0}
\end{center}
\caption{Dictionary between different notations
for the $B\to K^*\mu^+\mu^-$ observables
and Wilson coefficients they are most sensitive to (the sensitivity to $C_7^{(\prime)}$ is only present at low $q^2$).}
\label{tab:BKsllobs}
\end{table}

The angular distribution of the exclusive $\bar B\to \bar K^{*0}(\to K^-\pi^+)\mu^+\mu^-$ decay gives access to many observables potentially sensitive to NP~\cite{Kruger:2005ep,Lunghi:2006hc,Egede:2008uy,Altmannshofer:2008dz,Alok:2009tz,Lunghi:2010tr,Alok:2010zd,Bobeth:2010wg,Alok:2011gv,Bobeth:2011gi,Bobeth:2008ij}. By means of its charge conjugated mode $B\to K^{*0}(\to K^+\pi^-)\mu^+\mu^-$, which can be distinguished from the former simply by the meson charges, this decay allows a straightforward measurement of CP asymmetries.

Neglecting scalar operator contributions (which are strongly constrained by $B_s\to\mu^+\mu^-$) and lepton mass effects (which is a very good approximation for electrons and muons even if effects from collinear QED logarithms are taken into account~\cite{Huber:2005ig}),
the full set of observables accessible in the angular distribution of the decay and its CP-conjugate is given by 9+9 angular coefficients $I_i(q^2)$ and $\bar I_i(q^2)$, which are functions of the dilepton invariant mass $q^2$.
While the overall normalization of the angular coefficients is subject to considerable uncertainties, theoretically cleaner observables are obtained by normalizing them to the total invariant mass distribution. Furthermore, it makes sense to separate the observables into CP asymmetries $A_i$ and CP-averaged ones $S_i$. One thus arrives at \cite{Altmannshofer:2008dz}
\begin{equation}
 S_i = \left( I_i + \bar I_i \right) \bigg/ \frac{d(\Gamma+\bar\Gamma)}{dq^2} \,,
\qquad
 A_i = \left( I_i - \bar I_i \right) \bigg/ \frac{d(\Gamma+\bar\Gamma)}{dq^2}\,.
\label{eq:As}
\end{equation}
We will also consider observables integrated in a $q^2$ range, defined as
\begin{equation}
\langle S_i \rangle_{[a,b]}= \left(\int_a^b dq^2 \left( I_i + \bar I_i \right) \right)
\bigg/
\left(\int_a^b dq^2\frac{d(\Gamma+\bar\Gamma)}{dq^2}\right) \,,
\qquad
\label{eq:Asint}
\end{equation}
and analogously for $\langle A_i \rangle$.

For all observables one has to distinguish, both theoretically and experimentally, between the kinematical region where the dilepton invariant mass is below the charmonium resonances (low $q^2$ or large recoil region) and the region above (high $q^2$ or low recoil region). The intermediate region is of no interest to probe NP, as the $c\bar c$ resonances dominate the short distance rate by two orders of magnitude.

At low $q^2$, the observables are sensitive to all the Wilson coefficients $C_{7,9,10}^{(\prime)}$.
Among the CP asymmetries%
\footnote{We do not consider the CP asymmetries $A_{6s}^{V2s}$ and $A_8^V$ defined in \cite{Egede:2010zc}  since the former is suppressed by a small strong phase even beyond the SM and the latter is normalized to the quantity $I_8+\bar I_8$, which is zero at LO even beyond the SM and afflicted with considerable uncertainty.},
the most promising ones are then the T-odd CP asymmetries $A_7$, $A_8$ and $A_9$, which are not suppressed by small strong phases \cite{Bobeth:2008ij}.
At high $q^2$, the contributions of the magnetic penguin operators $C_7^{(\prime)}$ are suppressed, which in turn allows a cleaner sensitivity to the semi-leptonic operators. The CP asymmetries reduce to three independent ones, which are however T-even and therefore suppressed by small strong phases even beyond the SM \cite{Bobeth:2011gi}.
Among the CP-averaged angular coefficients, two have already been measured \cite{:2008ju,:2009zv,Aaltonen:2011cn,Aaltonen:2011ja,LHCb-CONF-2011-038}: the forward-backward asymmetry $A_\text{FB}$ and the $K^*$ longitudinal polarisation fraction $F_L$.
Recently, the CDF collaboration also published first bounds on $S_3$ and $A_9$ \cite{Aaltonen:2011ja}.
A promising observable in the early phase of LHC is the observable $S_5$ \cite{Bharucha:2010bb}.

Since different notations and conventions exist for the numerous $B\to K^*\ell^+\ell^-$ observables, in table~\ref{tab:BKsllobs} we provide a dictionary between the notation used in this work and a selection of other theory and experimental papers. It also lists the Wilson coefficients which, if modified by NP, would have the biggest impact on the observable in question. In the case of $C_7$ and $C_7'$, this sensitivity is only present at low $q^2$.

The main challenge in the theoretical prediction of the $B\to K^*\ell^+\ell^-$ observables is given on the one hand by the $B\to K^*$ form factors; on the other by non-factorisable effects\footnote{For a recent discussion of uncertainties in the low $q^2$ region see also \cite{Khodjamirian:2010vf}.}.
At low $q^2$, QCD factorisation can be used in the heavy quark limit, which reduces the number of independent form factors from 7 to 2 and allows a systematic calculation of non-factorizable corrections \cite{Beneke:2001at,Beneke:2004dp}. The remaining theoretical uncertainties then reside in phenomenological parameters like meson distribution amplitudes, in the form factors themselves, as well as in possible corrections of higher order in the ratio $\Lambda_\text{QCD}/m_b$.
Instead of using the two form factors in the heavy quark limit, we use the full set of seven form factors calculated by QCD sum rules on the light cone (LCSR), using the results of \cite{Ball:2004rg,Altmannshofer:2008dz}.
This approach has two advantages. First, using the full set of form factors
takes into account an important source of power suppressed corrections at low $q^2$. Second, the correlated uncertainties between the different form factors obtained from the LCSR calculation leads to a strongly reduced form factor uncertainty on observables involving ratios of form factors.

At high $q^2$, QCD factorization and LCSR methods are not applicable.
For the form factors, lacking predictions from lattice QCD, one currently has to rely on extrapolations of low-$q^2$  calculations, which introduce considerable uncertainty. For the estimation of non-factorizable corrections, an operator product expansion in powers of $1/\sqrt{q^2}$ can be used \cite{Grinstein:2004vb,Beylich:2011aq} and in Ref.~\cite{Beylich:2011aq} it has been argued that non-perturbative corrections {\em not} accounted for by the form factors are of the order of only a few percent.
We do take into account non-factorizable corrections proportional to form factors at $O(\alpha_s)$ both at low and high $q^2$ \cite{Beneke:2001at,Asatryan:2001zw,Beneke:2004dp,Seidel:2004jh,Greub:2008cy}

For our numerical analysis, a description of our treatment of theory uncertainties in the $B\to K^*\mu^+\mu^-$ observables is in order. For both high and low $q^2$, we take into account parametric uncertainties, varying the ratio $m_c/m_b$ from 0.25 to 0.33, the renormalization
scale from 4.0 to 5.6~GeV~\cite{Altmannshofer:2008dz} and the CKM angle $\gamma$ by $\pm11^\circ$~\cite{Charles:2004jd}.
At low $q^2$, as mentioned above, we make use of the LCSR calculation of all 7 form factors and vary the LCSR parameters as discussed in Ref.~\cite{Altmannshofer:2008dz}. To be conservative,
we add an additional real scale factor with an uncertainty of 10\% to each of the transversity amplitudes to account for possible additional power suppressed corrections.
The branching ratio is the only observable that is sensitive to the overall normalization of the form factors. Since LCSR only give predictions for the $B$ meson decay constant times a form factor, we add an additional relative uncertainty of twice the uncertainty of $f_B = (205\pm12)$~MeV~\cite{Laiho:2009eu} to the branching ratio. Unlike in \cite{Altmannshofer:2008dz}, we do not use the data on $B\to K^*\gamma$ to fix the form factor normalization, since we allow for NP also in $B\to K^*\gamma$.
We add all the individual uncertainties in quadrature.
At high $q^2$, we use the extrapolated form factors of Ref.~\cite{Bharucha:2010im}. In the Simplified Series Expansion used there, each form factor depends on two parameters fitted to the low-$q^2$ LCSR calculation. We estimate the form factor uncertainty by varying all 14 fit parameters separately, i.e. considering the uncertainties of the individual form factors as uncorrelated to each other, and add the resulting errors in quadrature. We consider this approach to be conservative. In view of the resulting sizable form factor uncertainties, at high $q^2$ we do not consider additional uncertainties due to power 
corrections or duality violation, which should amount to only a few percent \cite{Beylich:2011aq} and are therefore numerically irrelevant.

\begin{figure}[p]
\centering
\makebox[\textwidth]{%
\includegraphics[width=1.1\textwidth]{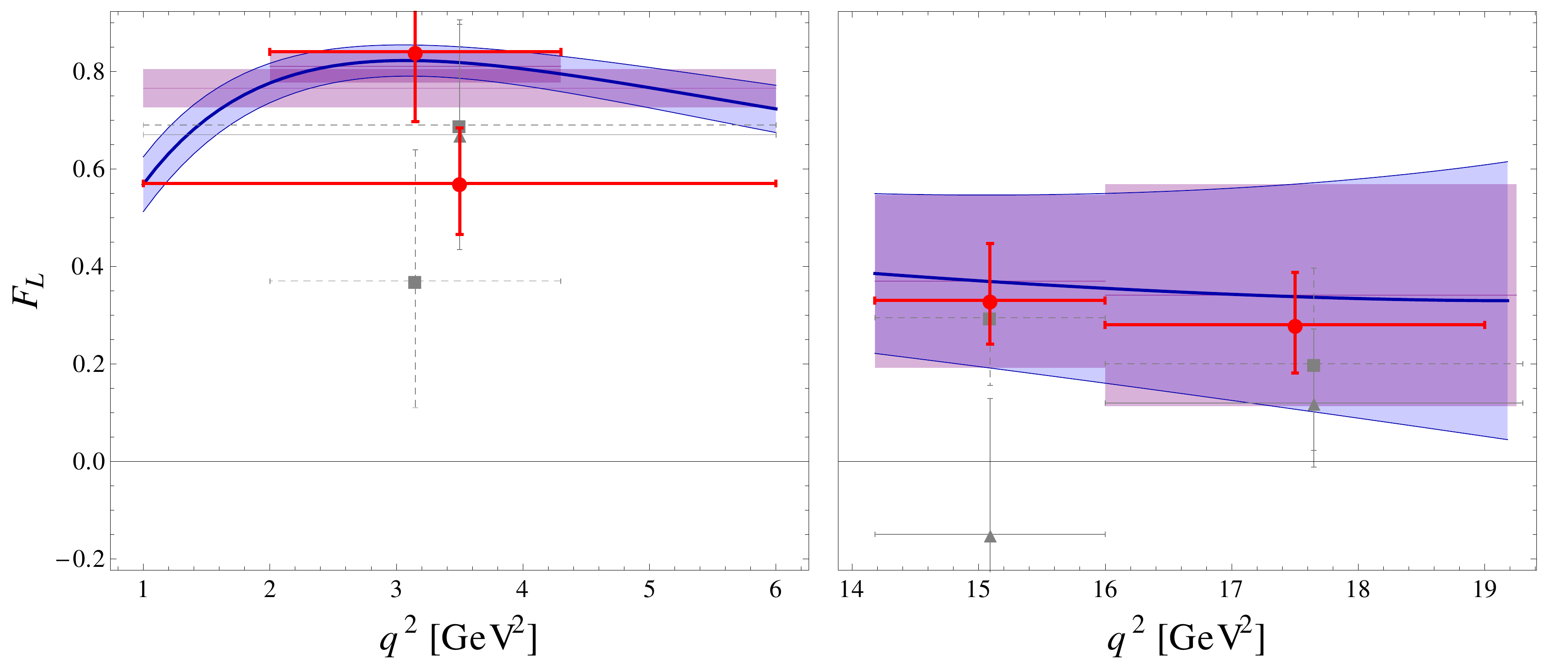}
}
\makebox[\textwidth]{%
\includegraphics[width=1.1\textwidth]{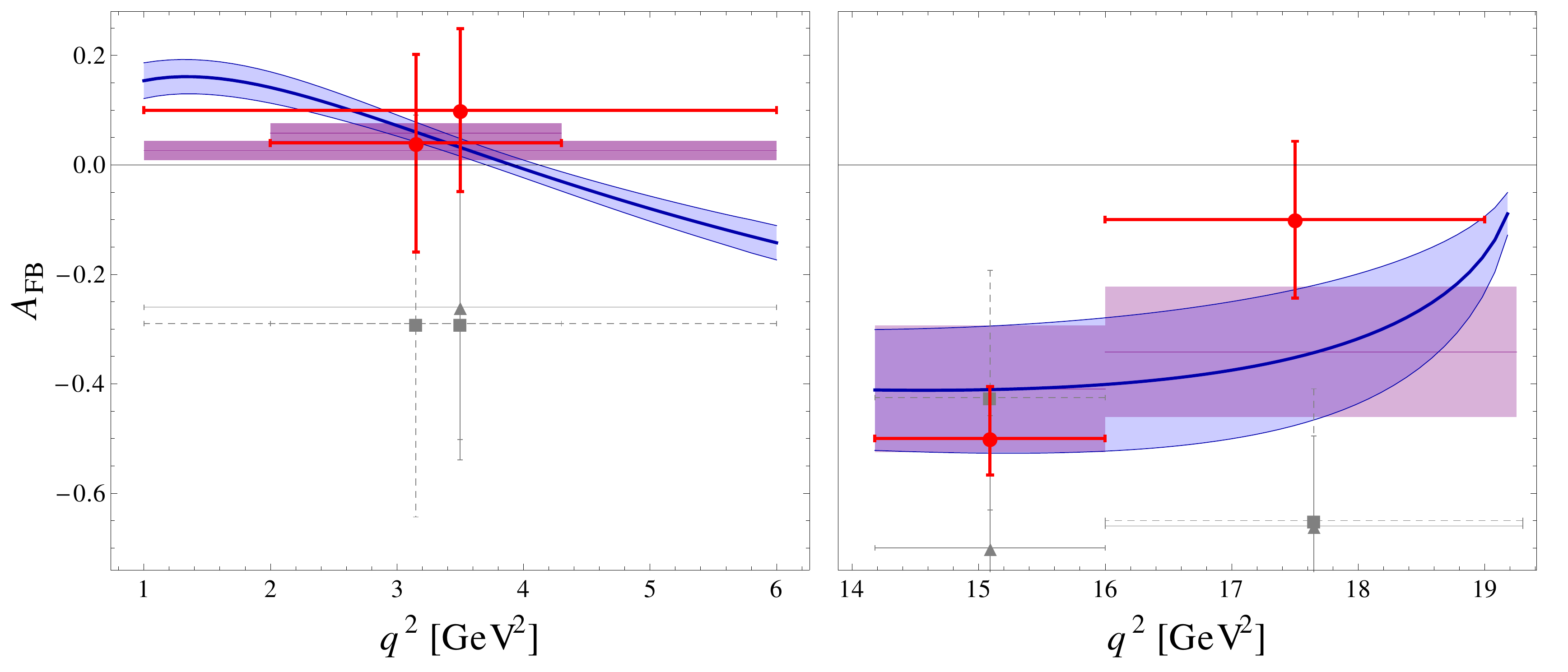}
}
\caption{Theory prediction (blue bands), binned theory prediction (purple horizontal bands) and experimental measurements from LHCb (red dots, thick error bars), CDF (gray squares, dashed error bars) and Belle (gray triangles, solid error bars) for $A_\text{FB}$ and $F_L$ in $B\to K^*\mu^+\mu^-$ for the two theoretically controllable $q^2$ regions.}
\label{fig:bksll}
\end{figure}

Figure~\ref{fig:bksll} shows the predictions for $F_L$ and $A_\text{FB}$ with our error estimates at low and high $q^2$ and compares them to the experimental data from Belle \cite{:2009zv}, CDF \cite{Aaltonen:2011qs} and LHCb \cite{LHCb-CONF-2011-038}. We do not show the data from BaBar \cite{:2008ju}, since they are given in large bins that include $q^2$ regions which are under poor theoretical control.
\section{Model-independent constraints on Wilson coefficients}\label{sec:modelind}

In a vast class of models beyond the SM, all NP effects in the observables listed in table~\ref{tab:obs} and discussed in section~\ref{sec:obs} are described by a modification of the Wilson coefficients $C_{7,8,9,10,S,P}^{(\prime)}$ at a matching scale, typically of the order of heavy particles contributing to the FCNC processes. For definiteness, we will consider in the following constraints on Wilson coefficients evaluated at the scale $\mu_h=160\,\text{GeV}$. The values at any other scale below the matching scale can be obtained straightforwardly using the renormalization group~\cite{Chetyrkin:1996vx,Gorbahn:2005sa}, see also appendix~\ref{sec:ceff}.

Up to subleading contributions, the coefficients $C_8^{(\prime)}$ enter the processes of interest only via renormalization group running by the mixing of the operators $\mathcal O_{7}^{(\prime)}$ and $\mathcal O_{8}^{(\prime)}$.
Therefore, we will not consider constraints on $C_8^{(\prime)}$ in the following and keep in mind that, in the presence of $C_8^{(\prime)}$, the constraints on  $C_7^{(\prime)\text{NP}}$ we present can be understood as constraints on the combination
$(C_7^{(\prime)\text{NP}} + 0.16 \,C_8^{(\prime)\text{NP}})$
at  $\mu_h$ (see appendix~\ref{sec:ceff}).
Among the observables we consider, the scalar and pseudoscalar coefficients $C_{S,P}^{(\prime)}$ can only affect the branching ratios of $B_s\to\mu^+\mu^-$ and $B_s\to\tau^+\tau^-$ in a significant way. Thus, we will disregard also these coefficients and instead use the constraints on the remaining Wilson coefficients to give a prediction for the maximum possible sizes of BR$(B_s\to\mu^+\mu^-)$ and BR$(B_s\to\tau^+\tau^-)$ {\em in the absence} of (pseudo)scalar operators.
We are thus left with the 6, potentially complex, Wilson coefficients $C_{7,9,10}^{(\prime)}$.

To obtain constraints on the Wilson coefficients, we construct a $\chi^2$ function, which is a function of Wilson coefficients $\vec C$ and contains the theory predictions for the observables $O_i^\text{th}$ and the experimental central values $O_i^\text{exp}$ as well as the corresponding uncertainties (which we assume to be Gaussian),
\begin{equation}
\chi^2(\vec C) =
\sum_i
\frac{\left( O_i^\text{exp} - O_i^\text{th}(\vec C) \right)^2}{(\sigma_i^\text{exp})^2+(\sigma_i^\text{th}(\vec C))^2}
\,.
\end{equation}
We write the theory uncertainty as a function of Wilson coefficients since, as discussed in section~\ref{sec:obs}, it is a relative error for some observables and for some others, such
as the $B\to K^*\mu^+\mu^-$ angular coefficients, can even be a non-trivial function of Wilson coefficients.

\begin{table}[t]
\addtolength{\arraycolsep}{10pt}
\renewcommand{\arraystretch}{1.3}
\begin{center}
\begin{tabular}{|l|cr|cr|}
\hline   
Observable & Experiment &  &  SM prediction &  \\  
\hline\hline
$10^4 \times$BR$(B \to X_s \gamma)$ & $3.55\pm 0.26$ & \cite{Asner:2010qj} & $3.15\pm 0.23$ & \cite{Misiak:2006zs} \\
\hline
$S_{K^*\gamma}$ & $-0.16 \pm 0.22$ & \cite{Asner:2010qj} & (-2.3$\pm$1.6)\% & \cite{Ball:2006eu} \\
\hline
$10^6 \times$BR$(B \to X_s \ell^+ \ell^-)_{[1,6]}$ & $1.63 \pm  0.50$ & \cite{Aubert:2004it,Iwasaki:2005sy} & $1.59 \pm 0.11$ & \cite{Huber:2005ig} \\
$10^7 \times$BR$(B \to X_s \ell^+ \ell^-)_{>14.4}$ & $4.3 \pm 1.2$ & \cite{Aubert:2004it,Iwasaki:2005sy} & $2.3 \pm 0.7$ & \cite{Hurth:2008jc} \\
\hline
$10^7 \times$BR$(B \to K^* \ell^+ \ell^-)_{[1,6]}$ & $1.71 \pm 0.22$ & \cite{Aaltonen:2011qs,:2009zv,LHCb-CONF-2011-038} & $2.28 \pm 0.63$ & \\
$10^7 \times$BR$(B \to K^* \ell^+ \ell^-)_{[14.18,16]}$ & $1.11 \pm 0.13$ & \cite{Aaltonen:2011qs,:2009zv,LHCb-CONF-2011-038} & $1.13 \pm 0.33$ & \\
$10^7 \times$BR$(B \to K^* \ell^+ \ell^-)_{[16,19]}$ & $1.35 \pm 0.15$ & \cite{Aaltonen:2011qs,:2009zv,LHCb-CONF-2011-038} & $1.34 \pm 0.51$ & \\
\hline
$\langle F_L\rangle(B \to K^* \ell^+ \ell^-)_{[1,6]}$ & $0.61\pm0.09$ & \cite{Aaltonen:2011ja,:2009zv,LHCb-CONF-2011-038} & $0.77\pm0.04$ & \\
$\langle F_L\rangle(B \to K^* \ell^+ \ell^-)_{[14.18,16]}$ & $0.28\pm0.09$ & \cite{Aaltonen:2011ja,:2009zv,LHCb-CONF-2011-038} & $0.37\pm0.17$ & \\
$\langle F_L\rangle(B \to K^* \ell^+ \ell^-)_{[16,19]}$ & $0.23\pm0.08$ & \cite{Aaltonen:2011ja,:2009zv,LHCb-CONF-2011-038} & $0.34\pm0.22$ & \\
\hline
$\langle A_{FB} \rangle(B \to K^* \ell^+ \ell^-)_{[1,6]}$ & $-0.04 \pm 0.12$ & \cite{Aaltonen:2011ja,:2009zv,LHCb-CONF-2011-038} & $0.03\pm0.02$ & \\
$\langle A_{FB} \rangle(B \to K^* \ell^+ \ell^-)_{[14.18,16]}$ & $-0.50\pm0.07$ & \cite{Aaltonen:2011ja,:2009zv,LHCb-CONF-2011-038} & $-0.41\pm0.11$ & \\
$\langle A_{FB} \rangle(B \to K^* \ell^+ \ell^-)_{[16,19]}$ & $-0.38\pm0.10$ & \cite{Aaltonen:2011ja,:2009zv,LHCb-CONF-2011-038} & $-0.35\pm0.11$ & \\
\hline
$\langle S_3 \rangle(B \to K^* \ell^+ \ell^-)_{[1,6]}$ & $0.27\pm0.56$ & \cite{Aaltonen:2011ja} & $(-0.3\pm1.1)\, 10^{-2}$ & \\
$\langle A_9 \rangle(B \to K^* \ell^+ \ell^-)_{[1,6]}$ & $0.09\pm0.39$ & \cite{Aaltonen:2011ja} & $(1.5 \pm 2.4)\, 10^{-4}$ & \\
\hline
\end{tabular}
\end{center}
\label{tab:obs} 
\caption{Experimental averages and SM predictions for the observables used in the fit.}
\end{table}

In table~\ref{tab:obs} we summarise the SM predictions as well as the experimental values
of the observables that we use in the $\chi^2$ function. To obtain the experimental values
we perform weighted averages of the available measurements, symmetrising the errors using
the prescription of ref.~\cite{D'Agostini:2004yu}

\subsection{Impact of observables on pairs of Wilson coefficients}\label{sec:2dim}

Using the $\chi^2$ function, we can obtain constraints on the real or imaginary parts of a pair of Wilson coefficients, or in the complex plane of a single Wilson coefficient. This approach, which is reminiscent of the CKM constraints in the $\bar\rho$-$\bar\eta$ plane, has the advantage that it allows to transparently show  the impact of individual observables on the constraints. A similar approach has been used e.g. in \cite{Bobeth:2008ij,Bobeth:2010wg,Bobeth:2011gi} for the SM Wilson coefficients and in \cite{DescotesGenon:2011yn} for $C_7$ vs. $C_7'$. Moreover, in some cases a pair of Wilson coefficients (or a single complex coefficient) captures already the dominant NP effect in certain scenarios, in which case such plots become particularly useful. For example, in the MSSM with MFV and flavour blind phases \cite{Altmannshofer:2008hc}, in effective SUSY with flavour blind phases \cite{Barbieri:2011vn} and in effective SUSY with a $U(2)^3$ symmetry \cite{Barbieri:2011ci,Barbieri:2011fc}, NP effects in $\Delta B=\Delta S=1$ processes arise almost exclusively through complex contributions to $C_7$ (and $C_8$). In MFV models with dominance of $Z$ penguins and without new sources of CP violation, only the real parts of $C_7$ and $C_{10}$ are relevant as will be discussed in section~\ref{sec:zpeng}.

\begin{figure}[p]
\centering
\includegraphics[width=\textwidth]{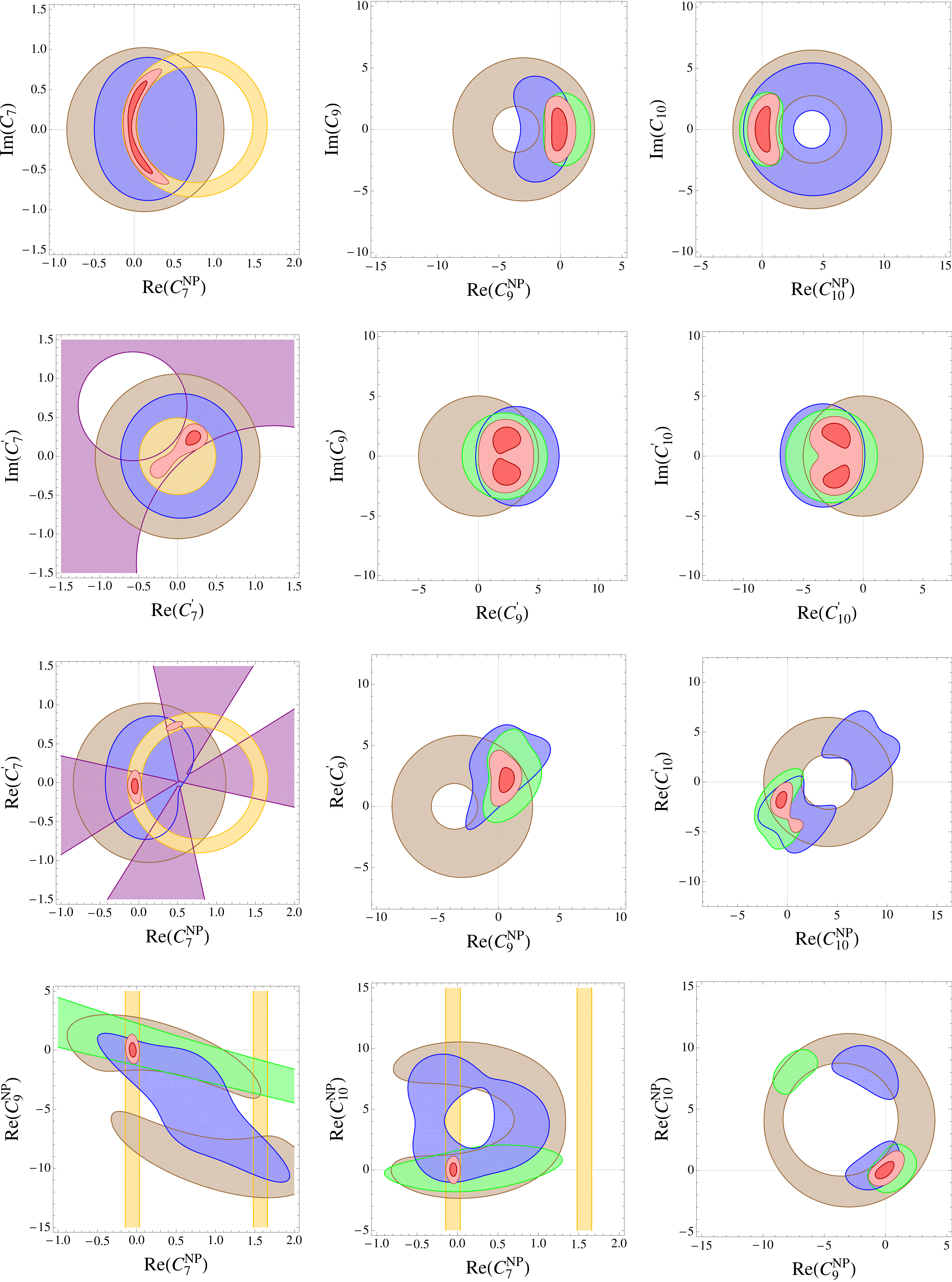}
\caption{Individual $2\sigma$ constraints on pairs of Wilson coefficients from  $B\to K^*\mu^+\mu^-$ at low $q^2$ (blue) and high $q^2$ (green), $B\to X_s\ell^+\ell^-$ (brown), BR($B\to X_s\gamma$) (yellow), $B\to K^*\gamma$ (purple) and combined 1 and $2\sigma$ constraints (red).}
\label{fig:bandplots}
\end{figure}

The resulting plots are shown in figure~\ref{fig:bandplots}. In these plots, the dark and light red regions show the 1 and $2\sigma$ best fit regions (contours of $\chi^2_\text{tot}-\chi^2_\text{tot,min}=1$ or 4), while the shaded regions in different colours show the $2\sigma$ allowed regions (contours of $\chi^2-\chi^2_\text{min}=4$) from $B\to K^*\mu^+\mu^-$ at low $q^2$ (blue), $B\to K^*\mu^+\mu^-$ at high $q^2$ (green), $B\to X_s\ell^+\ell^-$ (brown), BR($B\to X_s\gamma$) (yellow) and $B\to K^*\gamma$ (purple). We only show the observables that give relevant constraints.

We make several observations.
\begin{itemize}
\item At the 95\% C.L., all best fit regions are compatible with the SM.
\item In the complex $C_7$ plane, which is relevant e.g. for the models with flavour blind phases mentioned above, the inclusive and exclusive $b\to s\ell^+\ell^-$ observables -- in particular the measurement of BR$(B \to X_s \ell^+ \ell^-)$ at low $q^2$~\cite{Gambino:2004mv} and the LHCb measurement of $A_{\rm FB}$ at low $q^2$ -- exclude a sign-flip in the low-energy $C_7^\text{eff}$ that would be allowed by BR$(B\to X_s\gamma)$ and that was favoured by Belle data on $A_{\rm FB}$~\cite{:2009zv}.
We observe that an imaginary part of $C_7$ as large as $|{\rm Im}(C_7)| \lesssim 0.7$ is still allowed in this scenario.
\item In the presence of $C_7'$, the current data on the time-dependent CP asymmetry in $B\to K^*\gamma$ gives already an important constraint.
\item Similar to the complex $C_7$ plane, in the Re$(C_7)$--Re$(C_7^\prime)$ plane a sign-flip
in $C_7^\text{eff}$ is excluded by the inclusive and exclusive $b\to s\ell^+\ell^-$ observables. Imposing also the constraint from $S_{K^* \gamma}$ leaves only two disjoint regions at 95\% C.L.: one around the SM point and a second, less favoured one with a large Re$(C_7^\prime) \simeq 0.5$.
\item In the complex $C_9$ and $C_{10}$ planes, sign flips of the real parts are excluded
but imaginary parts as large as $|{\rm Im}(C_9)|, |{\rm Im}(C_{10})| \lesssim 3$ are still allowed.
\item The slight preference towards non-SM values of the Wilson coefficients in the complex $C_9^\prime$ and $C_{10}^\prime$ planes results from the tension between SM prediction and experimental data on BR$(B \to X_s \ell^+ \ell^-)$ in the high $q^2$ region as well as the
tension between SM and experimental data on $F_L$ in $B\to K^*\mu^+\mu^-$ at low $q^2$.
The tension in $F_L$ is also the reason for the slight preference for a non-SM value in the complex $C_7^\prime$ plane.
\item High-$q^2$ data on $B\to K^*\ell^+\ell^-$ are competitive with and coplementary to the low-$q^2$ ones.
In particular, they are crucial to exclude a sign flip in Re$(C_{10})$ as well as simultaneous
sign flips in $C_{10}$ and $C_{10}'$ or $C_{7}$ and $C_{9}$.
\item In the Re$(C_9)$--Re$(C_{10})$ plane, a simultaneous sign flip in $C_{9}$ and $C_{10}$ would be allowed by high $q^2$ data on $B\to K^*\ell^+\ell^-$ but is excluded by the new $A_{\rm FB}$ measurement at low $q^2$ from LHCb.
\end{itemize}

\noindent
We stress that the above conclusions hold if one allows only the two quantities shown to be non-zero. In many NP models, several of the Wilson coefficients will deviate from the SM, which renders some or all of these constraints ineffective. The more general case calls for a {\em global} fit of all Wilson coefficients.

\subsection{Global fit of Wilson coefficients}\label{sec:globalfit}

While the constraints discussed above are useful to display the constraining power of individual observables, they are not suited to put constraints on Wilson coefficients in models where more than two real or more than one complex coefficient is relevant, since cancellations can easily occur that render some of the constraints ineffective. Therefore, we now present constraints on Wilson coefficients varying not only 2 but all (or a subset) of the 6 complex coefficients $C_{7,9,10}^{(\prime)}$.
To cope with the large dimensionality of the parameter space, we perform a Markov Chain Monte
Carlo (MCMC) analysis. Details on the statistical approach are given in appendix~\ref{sec:stat}.

In concrete NP models, the contributions to the 6 Wilson coefficients $C_{7,9,10}^{(\prime)}$
are typically highly correlated and not all of them receive NP contributions. In addition to
a completely generic case, we will therefore consider several restricted scenarios, that are each representative for a vast class of models:
\begin{enumerate}
\item \textbf{Real left-handed currents}, $C_i\in\mathbbm{R}$, $C_i'=0$. This is realised e.g. in models with MFV in the definition of \cite{D'Ambrosio:2002ex,Buras:2000dm}, i.e. no CP violation beyond the CKM phase.
\item \textbf{Complex left-handed currents}, $C_i\in\mathbbm{C}$, $C_i'=0$. This is realised e.g. in models with MFV and flavour-blind phases.
\item \textbf{Complex right-handed currents}, $C_i'\in\mathbbm{C}$, $C_i=0$.
\item \textbf{Generic NP}, $C_i\in\mathbbm{C}$, $C_i'\in\mathbbm{C}$.
\end{enumerate}

\noindent
We remark that the results of the MFV setup analysed in~\cite{Hurth:2008jc}
are recovered as a limiting case of our scenario 1 when the flavour mixing angles
are taken to be CKM-like~\footnote{In principle, in MFV, we should also account
for the constraints arising from $b\to d$ and $s\to d$ transitions, such as
BR$(K_L\to \mu^+\mu^-)$, BR$(K\to \pi\ell^+\ell^-)$ or BR$(B \to X_d \gamma)$~\cite{Crivellin:2011ba}. In practise,
the latter turn out to be less stringent compared to the constraints from $b\to s$
transitions.}.

\subsubsection{Real left-handed currents}\label{sec:rLHC}

\begin{figure}[tbp]
\centering
\includegraphics[width=\textwidth]{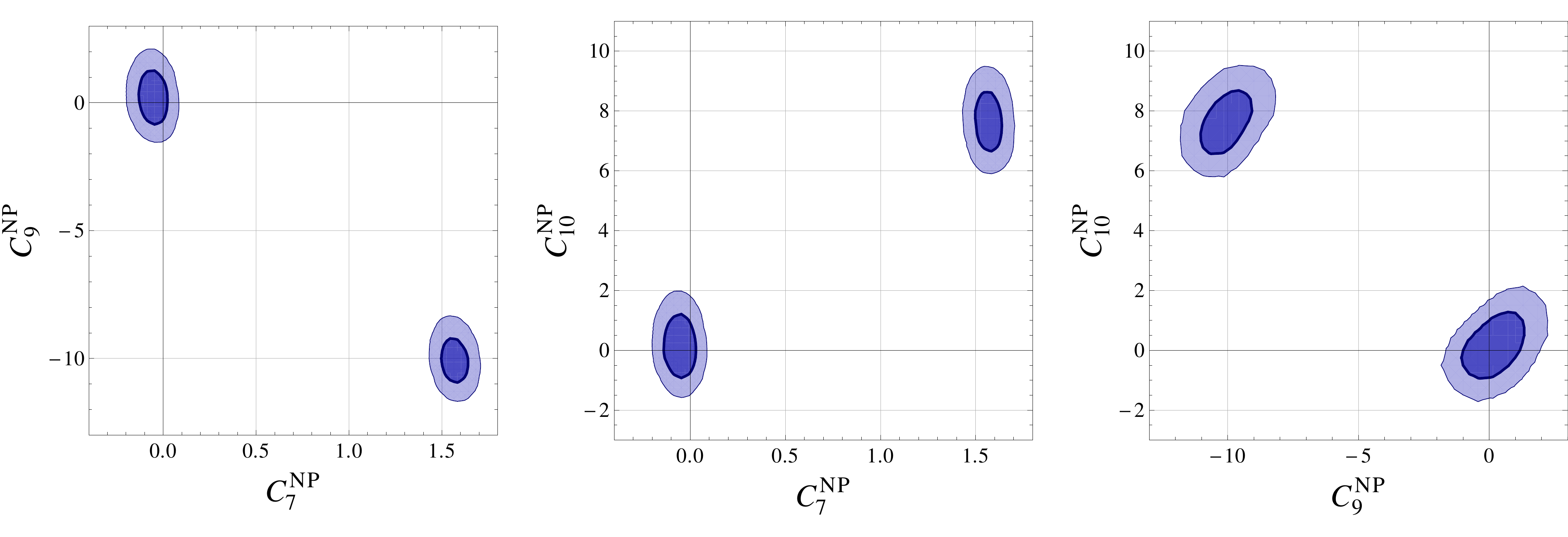}
\caption{Constraints between Wilson coefficients in the scenario with real left-handed currents. Shown are 68\% and 95\% C.L. regions.}
\label{fig:AllplotsMFV}
\end{figure}

In this scenario, there are only 3 free parameters, the real Wilson coefficients $C_{7,9,10}$. Fig.~\ref{fig:AllplotsMFV} shows the 68\% and 95\% confidence regions in the planes of two Wilson coefficients\footnote{Here and in the following, by ``C.L.'' we mean Bayesian confidence regions, i.e. regions containing 68\% or 95\% of the Markov chain points.}. While departures from the SM point are already quite strongly restricted, a crucial feature is the presence of a second allowed region, characterised by a simultaneous sign flip of $C_{10}$ as well as of the effective low-energy coefficients $C_7^\text{eff}$ and $C_9^\text{eff}$. A sign flip of only $C_9^\text{eff}$ and $C_7^\text{eff}$ or $C_{10}$ and $C_7^\text{eff}$ is excluded mainly by high-$q^2$ $B\to K^*\mu^+\mu^-$ data, as was discussed already in \cite{Bobeth:2010wg}.
Interestingly, a sign flip of only $C_9^{\rm eff}$ and $C_{10}$ is now excluded as well by the precise $A_{\rm FB}$ measurement at low $q^2$ (see section~\ref{sec:2dim}). 
We remark that an overall sign flip of all Wilson coefficients cannot be excluded by low-energy data alone, since all observables involve squared amplitudes or interference terms, which are invariant under such a sign flip. On the other hand, a NP model which generates effects in $C_7$, $C_9$ and $C_{10}$ which are each twice as large as their SM contributions seems highly unlikely. In the region with SM-like signs, we obtain the constraints
\begin{align}
C_7^\text{NP}&\in [-0.15,0.03]
\,,&
C_9^\text{NP}&\in [-1.1,1.6]
\,,&
C_{10}^\text{NP}&\in [-1.2,1.6]
\,,
\end{align}
at 95\% C.L.
These constraints can be translated into bounds on an effective NP scale $\Lambda$ that suppresses NP contributions to the corresponding higher dimensional operators in the effective Hamiltonian
\begin{equation}
{\cal H}_{\eff}^{\rm NP} = \frac{c_7^{bs}}{\Lambda^2_7} \mathcal O_7 + \frac{c_9^{bs}}{\Lambda^2_9} \mathcal O_9 + \frac{c_{10}^{bs}}{\Lambda^2_{10}} \mathcal O_{10} ~~ + \textnormal{h.c.} ~.
\end{equation}
Assuming the coefficients $c_i^{bs}$ to be 1, and using 95\% C.L. bounds on the absolute values of the Wilson coefficients we obtain
\begin{equation}
 \Lambda_7 > 55~\textnormal{TeV} ~,~~~ \Lambda_9 > 20~\textnormal{TeV} ~,~~~ \Lambda_{10} > 21~\textnormal{TeV} ~.
\end{equation}
These bounds on the effective NP scale are still weaker than the ones that can be obtained from considering dimension 6 operators that contribute to $B_s$ mixing~\cite{Isidori:2010kg}. 

\subsubsection{Complex left-handed currents}\label{sec:cMFV}

\begin{figure}[tbp]
\centering
\includegraphics[width=\textwidth]{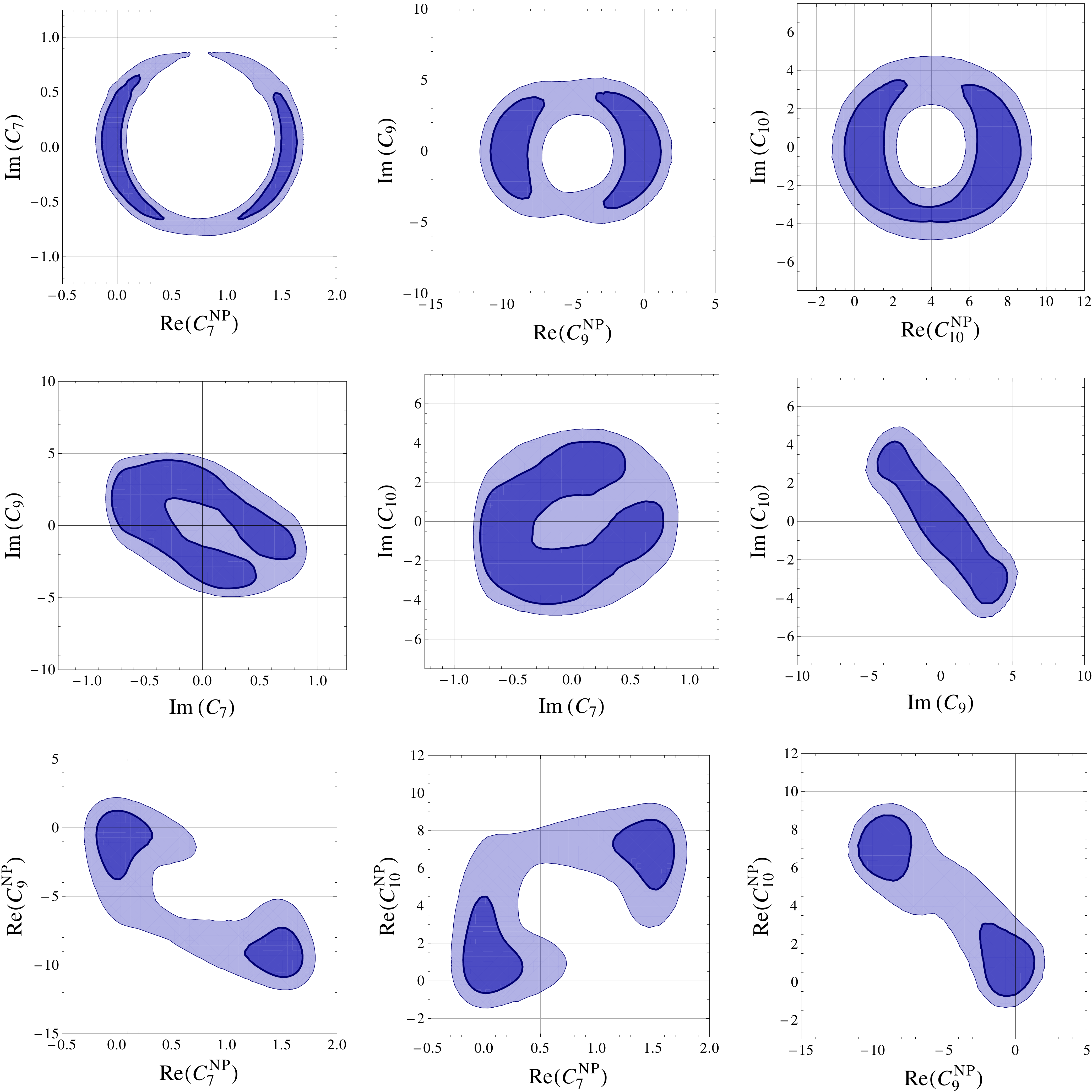}
\caption{Constraints between Wilson coefficients in the scenario with complex left-handed currents. Shown are 68\% and 95\% C.L. regions.}
\label{fig:AllplotscMFV}
\end{figure}

Complex contributions to the Wilson coefficients $C_{7,9,10}$ with vanishing $C_i'$ are predicted e.g. in models with MFV and flavour-blind phases. Fig.~\ref{fig:AllplotscMFV} shows the 68\% and 95\% confidence regions in the planes of two Wilson coefficients, omitting plots lacking a correlation. At 68\%~C.L., we observe two solutions for the real parts of the Wilson coefficients, corresponding to a SM-like case and a case with a simultaneous sign flip in the low-energy values of $C_{10}$, $C_{7}^\text{eff}$ and $C_{9}^\text{eff}$. However, the room for NP is much larger than in the case without non-standard CP violation since the increased number of free parameters allows for compensations among different
contributions.
In particular, sizable imaginary parts are allowed for all three coefficients
and this implies, in turn, potentially large effects in CP violating observables, as will be quantified in sec.~\ref{sec:predglobal2}.
A strong anti-correlation can be observed between the imaginary parts of $C_9$ and $C_{10}$,
driven mostly by the forward-backward asymmetry in $B\to K^*\mu^+\mu^-$ at high $q^2$.

\subsubsection{Complex right-handed currents}\label{sec:RHcurrents}

\begin{figure}[tbp]
\centering
\includegraphics[width=\textwidth]{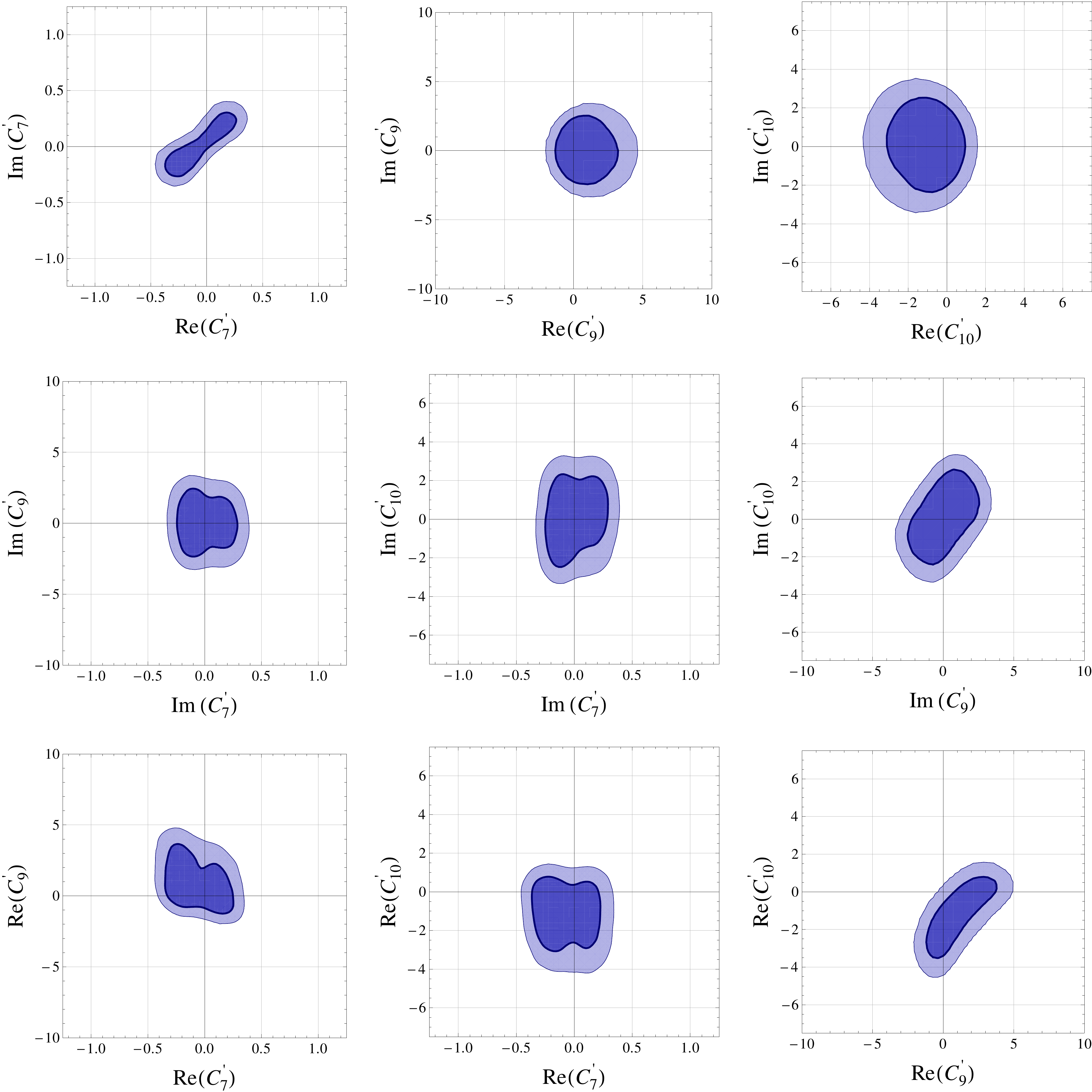}
\caption{Constraints between Wilson coefficients in the scenario with complex right-handed currents. Shown are 68\% and 95\% C.L. regions.}
\label{fig:Allplots_cRHC}
\end{figure}

Fig.~\ref{fig:Allplots_cRHC} shows the 68\% and 95\% confidence regions in the planes 
of two Wilson coefficients, omitting plots lacking a correlation. 
One of the most prominent differences we can observe compared to the case of left-handed curents,
is the absence of two solutions for the real and imaginary parts of the Wilson coefficients. 
This can be mainly traced back to the lack of interference between right-handed currents with the SM 
contributions in inclusive decays, which forbids sign flips in the low-energy values of the 
Wilson coefficients.
Also in this scenario the NP room for the imaginary parts of $C_7'$, $C_9'$ and
$C_{10}'$ is quite sizable and this will induce large effects in CP violating
observables, as discussed in sec.~\ref{sec:predglobal2}.
However, some of the correlations among low energy observables turn out to be different
compared to the case of left-handed current, providing a tool to distinguish the two scenarios.

\subsubsection{Generic NP}\label{sec:genNP}

\begin{figure}[tbp]
\centering
\includegraphics[width=\textwidth]{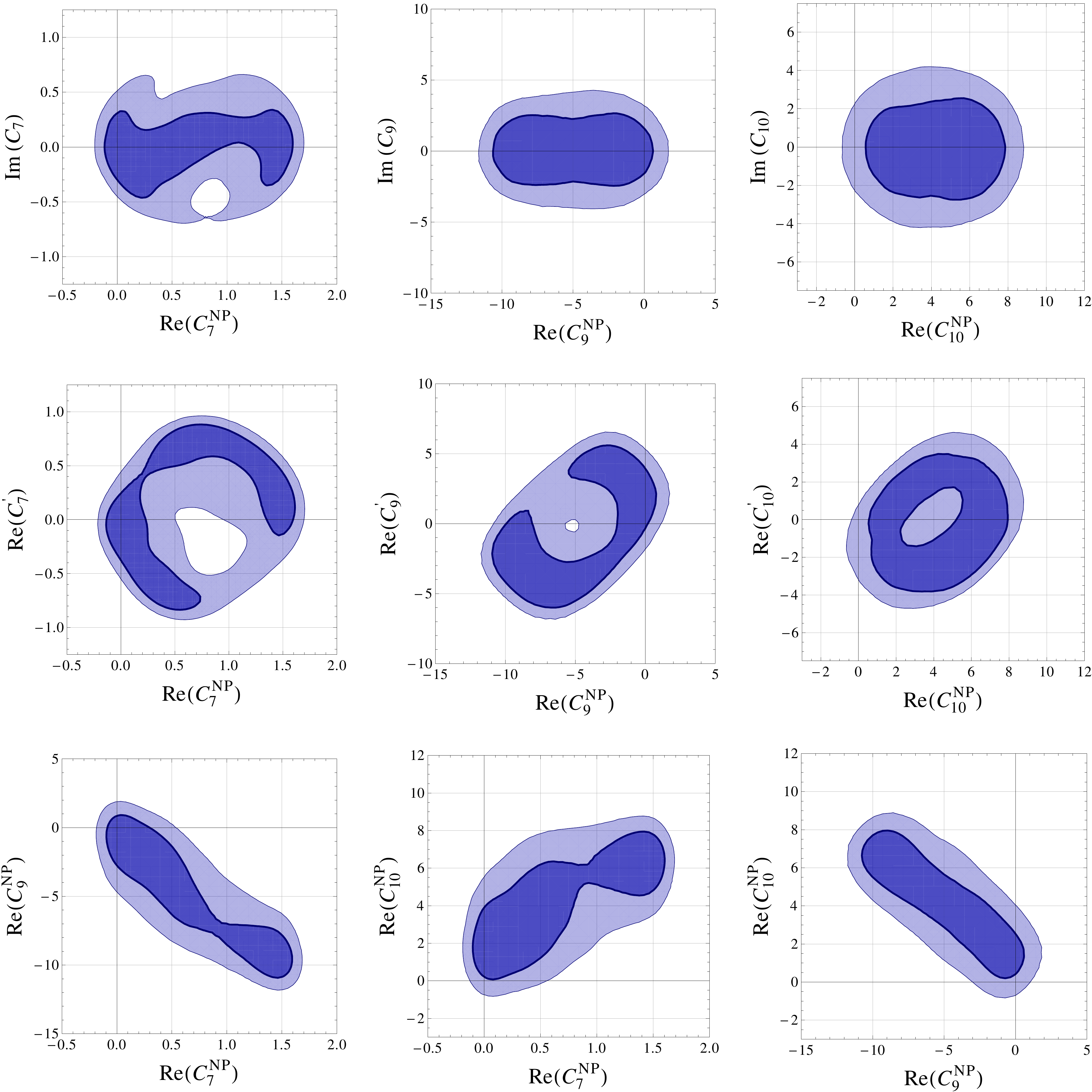}
\caption{Constraints between Wilson coefficients in the case of generic NP.
Shown are 68\% and 95\% C.L. regions.}
\label{fig:AllplotsgenNP}
\end{figure}

In this fit, we allowed all six Wilson coefficients $C_{7,9,10}^{(\prime)}$ to assume arbitrary complex values. Fig.~\ref{fig:AllplotsgenNP} shows the 68\% and 95\% confidence regions in the planes of two Wilson coefficients, omitting plots lacking a correlation. Sizable effects are not ruled out in any of the Wilson coefficients. However, some observations can be made, which are, by means of the generality of this fit, valid
{\em for any theory} beyond the SM.
\begin{itemize}
\item For $C_7$, $C_9$ and $C_{10}$ there is little room left for constructive interference of real NP contributions with the SM. Concretely, we find at 95\% C.L.
\begin{align}
\text{Re}\,C_7^\text{NP}&> -0.01
\,,&
\text{Re}\,C_9^\text{NP}&< 0.2
\,,&
\text{Re}\,C_{10}^\text{NP}&> 0.5
\,.
\end{align}
The slight preference for non-standard $C_{10}$ is again driven by the tension between SM
and experiment in BR$(B \to X_s \ell^+ \ell^-)$ at high $q^2$ and $F_L(B\to K^*\mu^+\mu^-)$ at low $q^2$.
\item a sizable negative real NP contribution to $C_9$ requires comparably large (with respect to the SM) {\em positive} contributions to $C_{10}$ and $C_7$.
\item Large imaginary parts for all coefficients and large chirality-flipped coefficients
are still allowed.
\end{itemize}
The last point highlights the importance of measuring observables sensitive to right-handed currents and to CP violation, such as the $B\to K^*\mu^+\mu^-$ observables $A_{7,8,9}$ and $S_3$.

\subsubsection{Predictions for $B_s\to\mu^+\mu^-$ and $B_s \to \tau^+\tau^-$}\label{sec:predglobal1}

\begin{table}[t] \small
\addtolength{\arraycolsep}{10pt}
\renewcommand{\arraystretch}{1.6}
\begin{center}
\makebox[\textwidth]{%
\begin{tabular}{|l|cccccc|}
\hline
Scenario & BR($\bsmm$) & BR($\bstt$) & $|\langle A_7\rangle_{[1,6]}|$ & $|\langle A_8\rangle_{[1,6]}|$ & $|\langle A_9\rangle_{[1,6]}|$ & $\langle S_3\rangle_{[1,6]}$  \\
\hline
Real LH & $[1.0,5.6] \times 10^{-9}$ & $[2,12] \times 10^{-7}$ & 0 & 0 & 0 & 0 \\
Complex LH & $[1.0,5.4] \times 10^{-9}$ & $[2,12] \times 10^{-7}$ & $<0.31$ & $<0.15$ & 0 & 0 \\
Complex RH & $<5.6\times 10^{-9}$ & $<12 \times 10^{-7}$ & $<0.22$ & $<0.17$ & $<0.12$ & $[-0.06,0.15]$ \\
Generic NP & $<5.5\times 10^{-9}$ & $<12 \times 10^{-7}$ & $<0.34$ & $<0.20$ & $<0.15$ & $[-0.11,0.18]$\\
\hline
LH $Z$ peng.& $[1.4,5.5] \times 10^{-9}$ & $[3,12] \times 10^{-7}$ & $<0.27$ & $<0.14$ & 0 & 0 \\
RH $Z$ peng.& $<3.8 \times 10^{-9}$ & $<8 \times 10^{-7}$ & $<0.22$ & $<0.18$ & $<0.12$ & $[-0.03,0.18]$ \\
Generic $Z$ p.& $<4.1 \times 10^{-9}$ & $<9 \times 10^{-7}$ & $<0.28$ & $<0.21$ & $<0.13$ & $[-0.07,0.19]$ \\
\hline
scalar current & $<1.1 \times 10^{-8}$ & $ <1.3 (2.3) \times 10^{-6}$ & 0 & 0 & 0 & 0\\
\hline
\end{tabular}
}
\end{center}
\caption{Predictions at 95\% C.L. for the branching ratios of $\bsmm$ and $\bstt$ and predictions
for low-$q^2$ angular observables in $B\to K^*\mu^+\mu^-$ (neglecting tiny SM effects
below the percent level) in all the scenarios. The scenarios ``Real LH'', ``Complex LH'', ``Complex RH'', ``Generic NP'', ``LH $Z$ peng.'', ``RH $Z$ peng.'', and ``Generic $Z$ p.'' correspond to the scenarios discussed in
sec.~\ref{sec:rLHC}, sec.~\ref{sec:cMFV}, sec.~\ref{sec:RHcurrents},
sec.~\ref{sec:genNP}, sec.~\ref{sec:LHZ}, sec.~\ref{sec:RHZ}, and
sec.~\ref{sec:LRHZ}, respectively, assuming negligible (pseudo)scalar currents.
In the scenario ``scalar current'' {\it only} scalar currents are considered.
The number quoted for $B_s\to\tau^+\tau^-$ in the ``scalar current''
scenario refers to the maximum value for its branching ratio in the case
of dominant scalar (pseudoscalar) currents.}
\label{tab:pred} 
\end{table} \normalsize

Due to its sensitivity to scalar currents, we did not include the branching ratios of $\bsmm$ and $\bstt$ in our fits. Instead, using the constraints on $C_{10}^{(\prime)}$, we can now give upper limits on the branching ratios in the considered scenarios assuming the scalar and pseudoscalar Wilson coefficients to be negligible. These bounds are useful since an observed violation of them would imply the presence of scalar currents.
The values we find at 95\% C.L. are listed in table~\ref{tab:pred}.
In the scenario with right-handed currents as well as in the case of generic NP, a significant suppression of the branching ratios is possible.
The upper bounds on BR($\bsmm$) in all cases are approximately a factor of 2 below the current experimental bound~(\ref{eq:bsmm_bound}) and correspond roughly to a 50\% 
enhancement of the branching ratio with respect to the SM. Due to the new measurement
of $B\to K^*\mu^+\mu^-$ angular observables, they are stronger than similar bounds
presented in the literature before \cite{Bobeth:2010wg}.

We stress that scalar current effects in $\bsmm$ could still enhance the branching ratio
over its current experimental bound.

Let us also mention that in the case of $B_d\to\mu^+\mu^-$, both the effects induced by the corresponding $b\to d$ Wilson coefficient $C_{10}^{(\prime)}$ and scalar currents can enhance the branching ratio
of $B_d\to\mu^+\mu^-$ over the experimental bound since the corresponding constraints 
from $b\to d\ell^+\ell^-$ processes are weaker. In addition, we remark that the ratio BR$(\bdmm)/\text{BR}(\bsmm)$ can significantly depart from the SM as well as
the MFV predictions $\text{BR}(\bdmm)/\text{BR}(\bsmm)\approx |V_{td}/V_{ts}|^2$ in both directions.

Finally, we discuss the allowed values for the branching ratio of $B_s\to\tau^+\tau^-$. From the general expressions of $\text{BR}(B_s\to\ell^+\ell^-)$ in presence of NP (see eq.~\ref{eq:BRbsmm})
one has
\begin{equation}
\frac{\text{BR}(\bstt)}{\text{BR}(\bsmm)} \simeq
\left(1 - \frac{4m^2_{\tau}}{m^2_{B_s}}\right)^{1/2}
\frac{m^{2}_{\tau}}{m^{2}_{\mu}}\times
\frac{\left(1-4m^2_{\tau}/m^2_{B_s}\right)|S|^2 + |P|^2}{|S|^2 + |P|^2}~,
\label{eq:bstt}
\end{equation}
where $S$ and $P$ have been defined in eq.~\ref{eq:SP}.

In the case where $C_{10}^{(\prime)}$ provides the dominant NP effects, one obtains
\begin{equation}
\frac{\text{BR}(\bstt)}{\text{BR}(\bsmm)} \simeq 212 ~,
\end{equation}
which implies, in particular, the SM prediction for the branching ratio of $B_s\to\tau^+\tau^-$
\begin{equation}
\text{BR}(B_s\to\tau^+\tau^-)_\text{SM}=(7.7\pm0.8)\times10^{-7} \,.
\end{equation}
In the case where (pseudo)scalar current effects dominate, BR$(B_s\to\mu^+\mu^-)$
can saturate the current experimental bound while for BR$(B_s\to\tau^+\tau^-)$ we get
\begin{equation}
\label{eq:bstt_vs_bsmm}
120\lesssim\frac{\text{BR}(B_s\to\tau^+\tau^-)}{\text{BR}(B_s\to\mu^+\mu^-)}
\lesssim 212\,,
\end{equation}
where the lower (upper) bound in eq.~\ref{eq:bstt_vs_bsmm} correspond to
the case where the scalar (pseudoscalar) contribution dominates.

Combining eq.~(\ref{eq:bsmm_SM2}) with eq.~(\ref{eq:bstt_vs_bsmm}), it turns out that BR$(B_s \to \tau^+ \tau^-) \lesssim 2 \times 10^{-6}$ (see also table~\ref{tab:pred}) and it has to be seen whether such values might be within the reach 
of LHCb. However, we stress that this upper bound relies on the assumption that 
the (pseudo)scalar Wilson coefficients $C^{(\prime)}_{S,P}$ are linearly proportional to the lepton Yukawa couplings, as discussed in sec.~\ref{sec:obs}. If we relax this assumption, as it might be the case in models 
like R-parity violating SUSY~\cite{Barbier:2004ez} or models with enhanced couplings with the third lepton 
generation (compared to the linear scaling assumed throughout this work), BR$(B_s \to \tau^+ \tau^-)$ could get in principle much larger values than $10^{-6}$.

\subsubsection{Predictions for $B\to K^*\mu^+\mu^-$}\label{sec:predglobal2}

\begin{figure}[tbp]
\centering
\includegraphics[width=\textwidth]{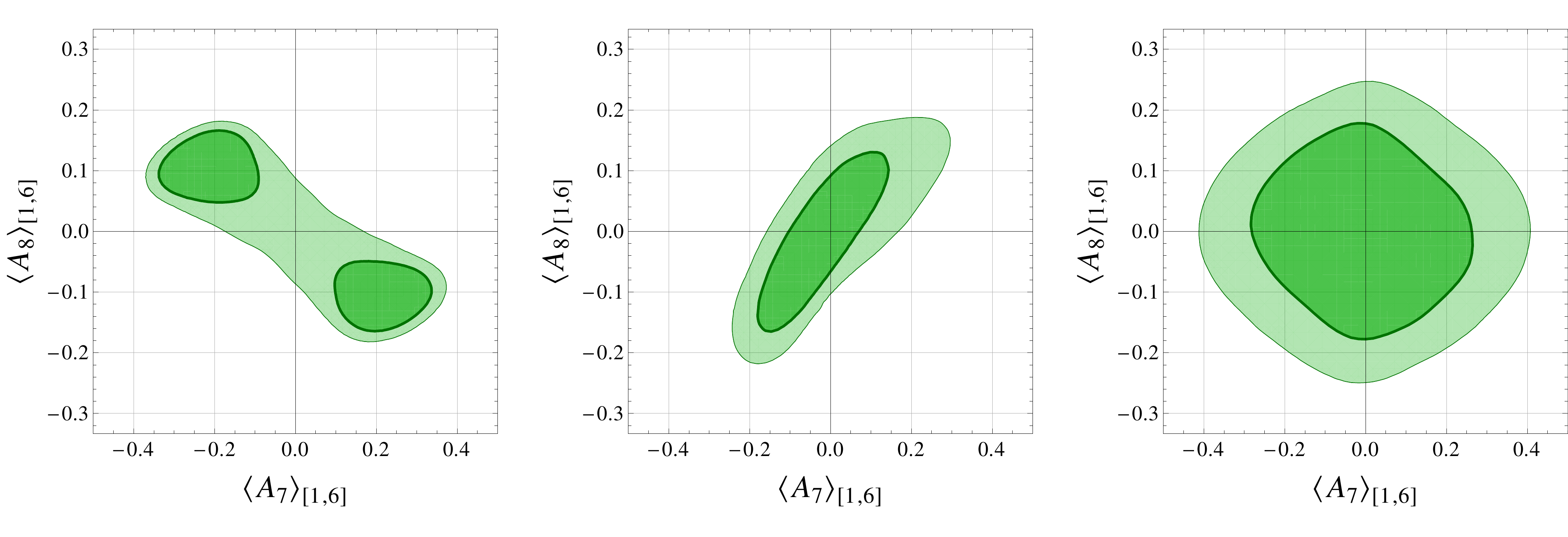}
\caption{Fit predictions for the low-$q^2$ CP asymmetries $\langle A_{7,8}\rangle$ in $B\to K^*\mu^+\mu^-$ in the case of complex left-handed currents (left), complex right-handed currents (centre) and generic NP (right). Shown are 68\% and 95\% C.L. regions.}
\label{fig:ObsPlots}
\end{figure}

Figure~\ref{fig:ObsPlots} shows the predictions for the T-odd $B\to K^*\mu^+\mu^-$ CP asymmetries $A_7$ and $A_8$ at low $q^2$ for the scenarios with complex left-handed currents, complex right-handed currents and for generic NP. 
In the absence of right-handed currents, one finds an anti-correlation between $A_7$ and $A_8$ which has already been found in models where only $C_7$ contributes \cite{Altmannshofer:2008dz,Barbieri:2011vn,Barbieri:2011fc} (see also \cite{Straub:2010ih}), but is shown here to hold under more general conditions.
At 68\% C.L., one finds a preference for non-standard CP asymmetries driven mostly by the tension between SM and experiment in $F_L(B\to K^*\mu^+\mu^-)$ at low $q^2$.
Similarly, in the absence of complex left-handed currents, one finds an opposite correlation.
In the generic case, there is no correlation at all.
Interestingly, in all three scenarios, large effects in both asymmetries are still allowed, with the numerical bounds listed in table~\ref{tab:pred}.
Future measurements of $A_7$ and $A_8$ at LHCb will thus be crucial to constrain the imaginary parts of the Wilson coefficients entering the $B\to K^*\mu^+\mu^-$ decay.

Also shown in table~\ref{tab:pred} are the predictions for the CP asymmetry $A_9$ and the CP-averaged angular coefficient $S_3$ at low $q^2$, both of which are tiny in the SM but can be sizable in presence of right-handed currents.
Indeed, both observables can assume values in excess of 10\% in the complex right-handed scenario and for generic NP.

We note that the CDF measurement of $S_3$ and $A_9$ shown in the last two rows of table~\ref{tab:obs} currently puts no significant constraints on NP, yet. Future measurements at LHCb with errors of the order of 0.1 will however put important constraints on CP-violating or CP-conserving right-handed currents.
\section{Analysis of flavour-changing $Z$ couplings}\label{sec:zpeng}

Tree level FCNC couplings of the $Z$ boson can appear in a number of NP scenarios.
Prominent examples are the SM with four non-sequential generations of quarks and models with an extra $U(1)$ symmetry~\cite{Langacker:2000ju,Langacker:2008yv}. Moreover, the contributions to the semi-leptonic operators are dominated
by $Z$ penguins, i.e. loop-induced modified $Z$ couplings\footnote{We remark that the distinction between $Z$ penguins and other contributions is in general gauge dependent. However, this gauge dependence is weak if $Z$ penguins dominate \cite{Buras:1998ed,Buchalla:2000sk,Altmannshofer:2009ma}.}, in many theories, e.g. in the MSSM~\cite{Lunghi:1999uk,Buchalla:2000sk}.
It is therefore interesting to consider the effects in a framework with modified  $\bar sbZ$ couplings, which can be  parametrised by the effective Lagrangian \cite{Buchalla:2000sk}
\begin{equation}
\mathcal L^{\bar sbZ}_{\rm eff} =
-\frac{G_F}{\sqrt{2}} \frac{e}{\pi^2}m_Z^2 c_w s_w V^*_{tb}V_{ts} \; Z^\mu
\left( Z_L \; \bar s \gamma_\mu P_L  b + Z_R \; \bar s \gamma_\mu P_R  b \right)~,
\label{eq:LbsZ}
\end{equation}
with $s_w =\sin\theta_w$ and $c_w =\cos\theta_w$. In this class of models one finds
\begin{align}
C_{10}^\text{NP} &= Z_L^\text{NP}
\,,&
C_{10}' &= Z_R
\,,\\
C_{9}^\text{NP} &= -Z_L^\text{NP}(1-4s_w^2)
\,,&
C_{9}' &= -Z_R(1-4s_w^2)
\,.
\end{align}
The contributions to $C_9^{(\prime)}$ are strongly suppressed by the small vector coupling
of the $Z$ to charged leptons $(1-4s_w^2)\approx0.08$.

The modified $\bar sbZ$ couplings also modify decays with a neutrino pair in the final state,
so one obtains correlations between $b\to s\nu\bar\nu$ and $b\to s \ell^+\ell^-$ observables. Writing the $b\to s\nu\bar\nu$ effective Hamiltonian as
\begin{gather} \label{eq:Heffnu}
{\cal H}_{\eff} = - \frac{4\,G_F}{\sqrt{2}}V_{tb}V_{ts}^*
\left(
C_L \mathcal O_L +C_R \mathcal O_R
\right) ~+~ {\rm h.c.} \,,
\\
\mathcal{O}_{L,R} =\frac{e^2}{8\pi^2}
(\bar{s} \gamma_{\mu} P_{L,R} b)( \bar{\nu} \gamma^\mu P_L \nu)\,,
\end{gather}
The effective Lagrangian (\ref{eq:LbsZ}) leads to
\begin{equation}
C_L = C_L^{\rm SM} + Z_L^\text{NP}
\,,\qquad
C_R = Z_R
\,,
\end{equation}
where $C_L^{\rm SM} = - 6.38 \pm 0.06$ \cite{Altmannshofer:2009ma}.

Finally, an effective tree level contribution to $B_s$-$\bar B_s$ mixing is generated by the exchange of a $Z$ with modified $\bar sbZ$ coupling. Its contribution to the mixing amplitude can be written at the scale $m_Z$ as~\cite{Altmannshofer:2009ma}
\begin{equation}
\frac{%
 \langle B_s | \mathcal H | \bar B_s \rangle^{\bar sbZ} }{%
 \langle B_s | \mathcal H | \bar B_s \rangle^{\rm SM} }
	= \frac{4\alpha s_w^2}{\pi S_0(x_t)} ( Z_L^2 -3.5 \,Z_L Z_R + Z_R^2 )~.
\label{eq:DMs-ZLR}
\end{equation}

\noindent
We consider three scenarios in the following:
\begin{itemize}
\item \textbf{\boldmath Left-handed modified $Z$ couplings}, $Z_L\in \mathbbm{C}$, $C_7^{\rm NP}\in\mathbbm{C}$, $Z_R=C_7'=0$,
\item \textbf{\boldmath Right-handed modified $Z$ couplings}, $Z_R\in \mathbbm{C}$, $C_7'\in\mathbbm{C}$, $Z_L^{\rm NP}=C_7^{\rm NP}=0$,
\item \textbf{\boldmath Generic modified $Z$ couplings}, $Z_{L,R}\in \mathbbm{C}$, $C_7^{(\prime)}\in\mathbbm{C}$,
\end{itemize}
allowing for non-standard CP violation in all cases. The generic case covers all NP models
where contributions to the semileptonic operators are dominated by $Z$ penguins. This includes
in particular the general MSSM~\cite{Lunghi:1999uk,Buchalla:2000sk}. The fitting procedure is as described in section~\ref{sec:globalfit}.

\subsection{Constraints on modified $Z$ couplings}

\begin{figure}[tbp]
\centering
\includegraphics[width=\textwidth]{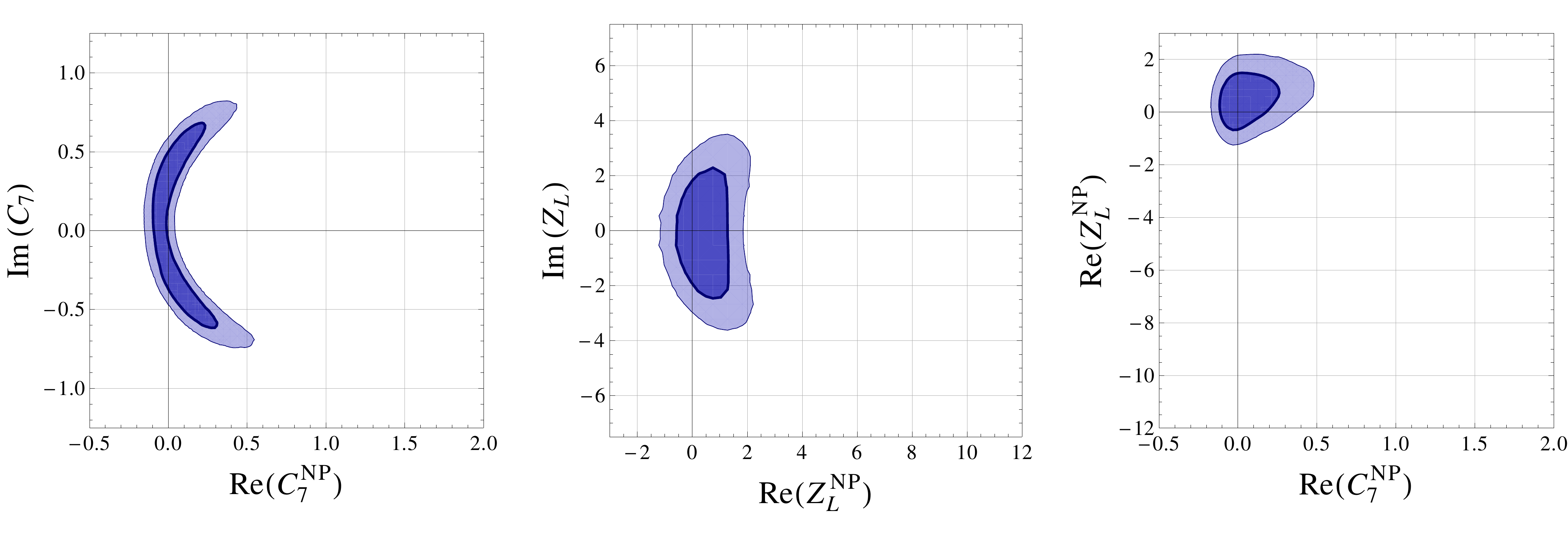}
\caption{Constraints on $C_7$ and the modified $Z$ coupling in the scenario with left-handed couplings only. Shown are 68\% and 95\% C.L. regions.}
\label{fig:AllplotsMFVZ}
\end{figure}

\subsubsection{Left-handed modified $Z$ couplings}\label{sec:LHZ}

Fig.~\ref{fig:AllplotsMFVZ} shows the 68\% and 95\% confidence regions in the complex planes
of $C_7$ and $Z_L$ as well as the correlation between the real parts of $C_7$ and $Z_L$.
The constraint in the $C_7$ plane is basically identical to the constraint in the absence of semileptonic operators shown in the upper left plot of figure~\ref{fig:bandplots}, which is
in contrast to the corresponding constraint in the presence of $C_9$ and $C_{10}$ shown in figure~\ref{fig:AllplotscMFV}, where large effects in Re$(C_7)$ (and in particular a sign flip) were allowed. This can be traced back to the suppression of $C_9$ compared to $C_{10}$ in the
$Z$ penguin scenario, making impossible a simultaneous sign flip of $C_7^\text{eff}$ and $C_9^\text{eff}$ at low energies, which would be required in particular to meet the constraint
from $A_\text{FB}(B\to K^*\mu^+\mu^-)$ at low $q^2$. For the same reason, the correlation of
the real parts of $C_7$ and $Z_L$ only shows one solution.

\begin{figure}[tbp]
\centering
\includegraphics[width=\textwidth]{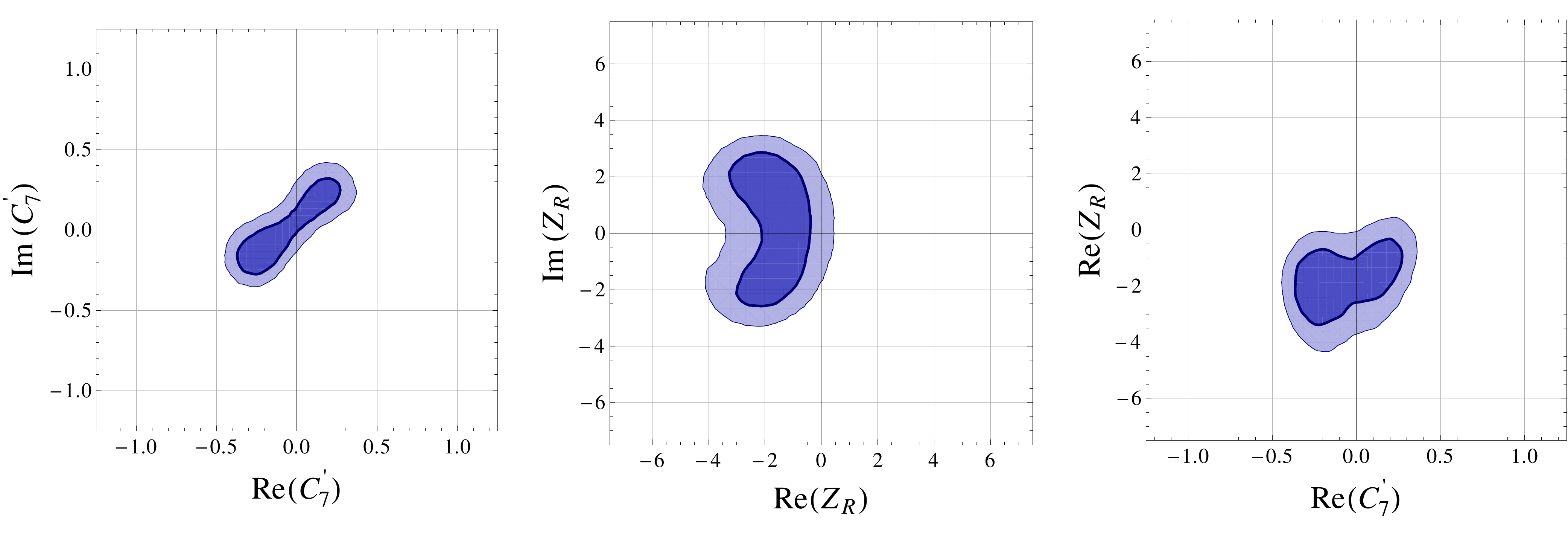}
\caption{Constraints on $C_7$ and the modified $Z$ coupling in the scenario with right-handed couplings only. Shown are 68\% and 95\% C.L. regions.}
\label{fig:AllplotsRHZ}
\end{figure}
\subsubsection{Right-handed modified $Z$ couplings}\label{sec:RHZ}

Fig.~\ref{fig:AllplotsRHZ} shows the 68\% and 95\% confidence regions in the complex planes
of $C_7'$ and $Z_R$ as well as the correlation between the real parts of $C_7'$ and $Z_R$.
The constraint in the $C_7'$ plane is very similar to the constraint in the absence of semileptonic operators shown in the corresponding plot of figure~\ref{fig:bandplots}.
The negative values preferred for the real part of the right-handed $Z$ coupling, i.e. for  Re$(C_{10}')$, arises from low- and high $q^2$ $B\to K^*\mu^+\mu^-$ data, as can be seen in the corresponding plot of fig.~\ref{fig:bandplots}.

\begin{figure}[tbp]
\centering
\includegraphics[width=0.66\textwidth]{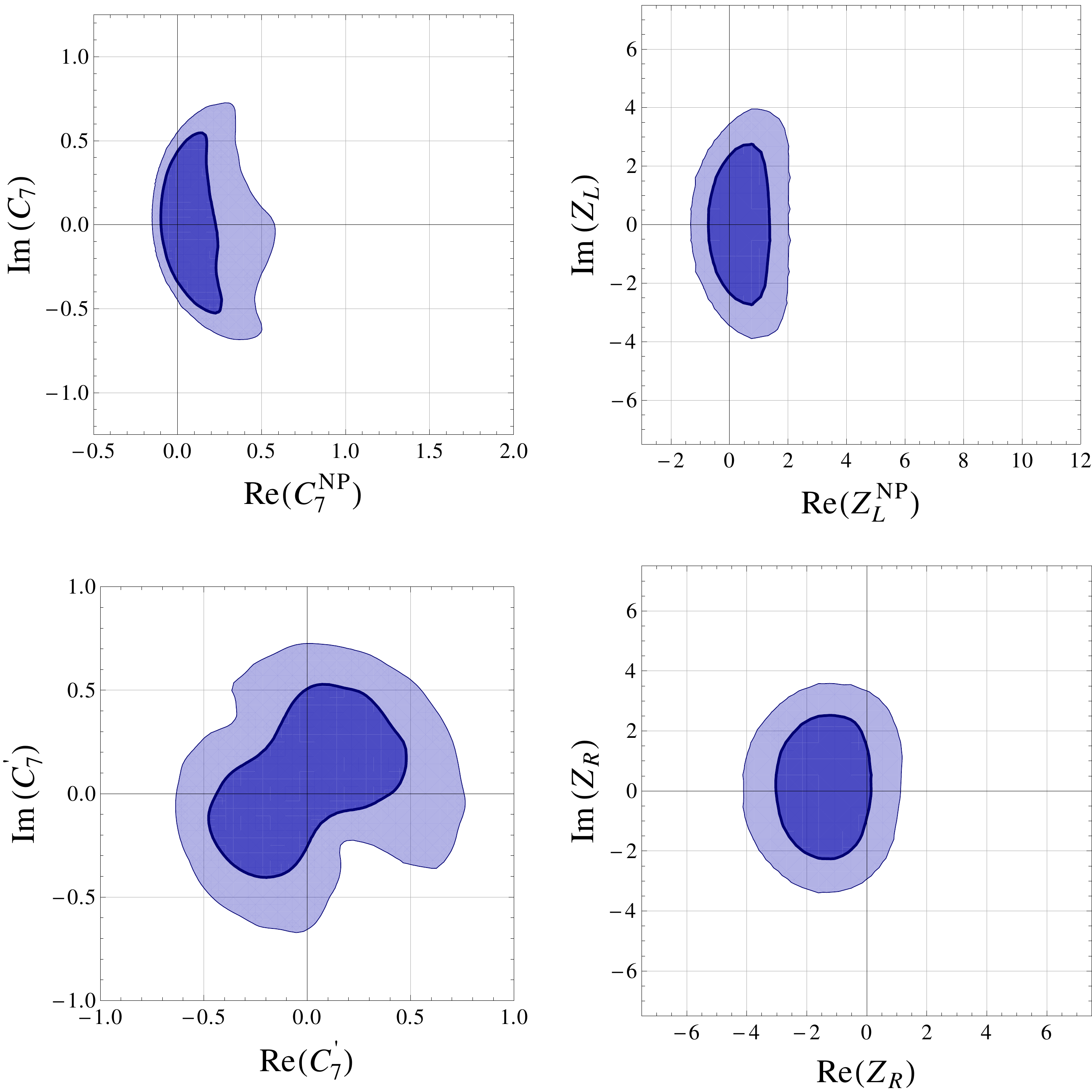}
\caption{Constraints on $C_7^{(\prime)}$ and the modified $Z$ couplings in the scenario with generic modified $Z$ couplings. Shown are 68\% and 95\% C.L. regions.}
\label{fig:AllplotsgenZ}
\end{figure}

\subsubsection{Generic modified $Z$ couplings}\label{sec:LRHZ}

Fig.~\ref{fig:AllplotsgenZ} shows the 68\% and 95\% confidence regions in the complex planes of the Wilson coefficients $C_7^{(\prime)}$ and the couplings $Z_{L,R}$ in the case of generic modified $Z$ penguins. While the room for NP is larger than in the more constrained previous cases, also in the generic there are no disjoint solutions for the Wilson coefficients. We are thus lead to conclude on a model-independent basis that if the NP contributions to semi-leptonic operators are dominated by $Z$ penguins, the real parts of the Wilson coefficients $C_{7,9,10}$ at low energies must have the same sign as in the SM.

\subsection{Fit predictions}

\subsubsection{Predictions for $B_s\to\mu^+\mu^-$, $B_s\to\tau^+\tau^-$  and $B\to K^*\mu^+\mu^-$}

\begin{figure}[tbp]
\centering
\includegraphics[width=\textwidth]{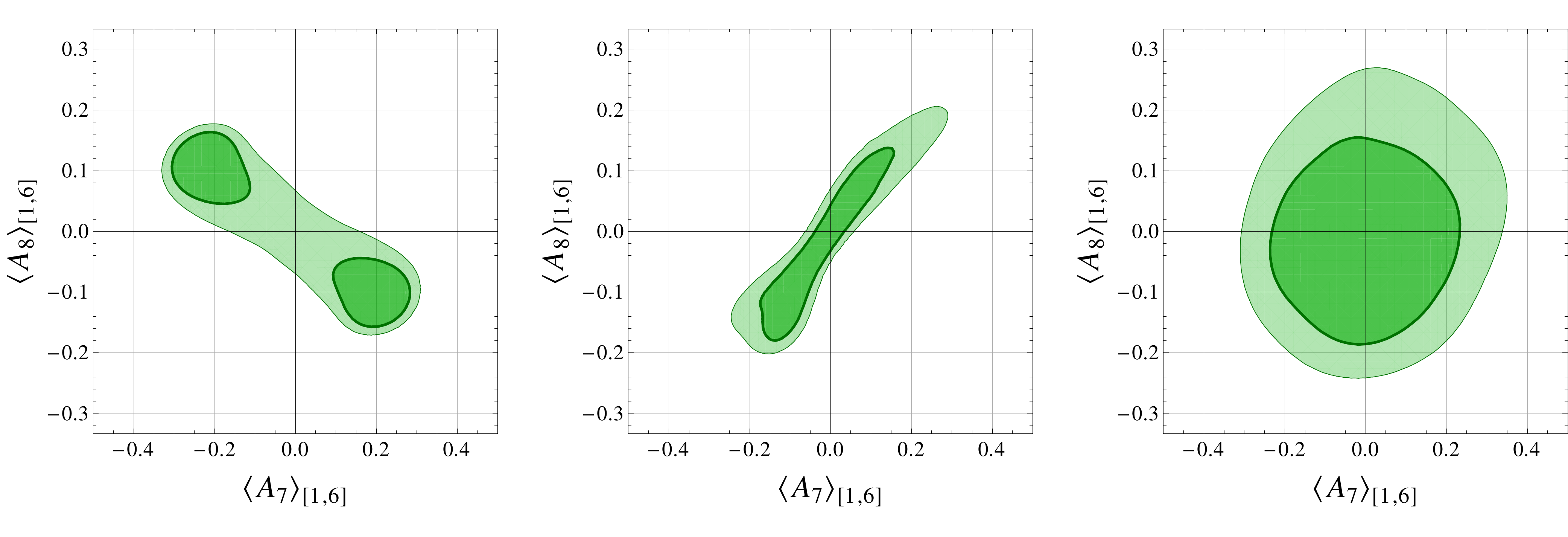}
\caption{Fit predictions for the low-$q^2$ CP asymmetries $\langle A_{7,8}\rangle$ in $B\to K^*\mu^+\mu^-$
for the scenario with left-handed (left), right-handed (centre) or generic (right) modified $Z$ couplings. Shown are 68\% and 95\% C.L. regions.}
\label{fig:A7A8Z}
\end{figure}

Analogously to section~\ref{sec:predglobal1}, we can give fit predictions for BR($B_s\to\mu^+\mu^-$) and BR($B_s\to\tau^+\tau^-$) in the absence of scalar currents and for $B\to K^*\mu^+\mu^-$ observables in the considered modified $Z$ coupling scenarios based on the constraints obtained in the global fit.
The allowed ranges are shown in table~\ref{tab:pred}.
In the generic case, the preference for smaller values of the $\bsmm$ and $\bstt$ branching ratios is due to the negative values preferred for Re$(Z_R)$ (cf. section~\ref{sec:RHZ}), i.e. for  Re$(C_{10}')$, which leads to a destructive interference with the SM in the decay amplitudes, see eq.~(\ref{eq:SP}).

\bigskip
Figure~\ref{fig:A7A8Z} shows the prediction for the $B\to K^*\mu^+\mu^-$ CP asymmetries $A_7$ and $A_8$ at low $q^2$ in all three scenarios.
The predictions are similar to the corresponding ones obtained for generic $C_{9,10}^{(\prime)}$ shown in figure~\ref{fig:ObsPlots}, so the comments made there apply here as well.
Also for the observables $S_3$ and $A_9$ we find predictions that are similar to the cases discussed in section~\ref{sec:predglobal2}. The results are summarised in table~\ref{tab:pred}.

\subsubsection{Predictions for $B_s$ mixing}

Since the real and imaginary parts of the left- and right-handed $Z$ couplings are constrained by $b\to s\ell^+\ell^-$ processes not to be significantly larger than the SM value of the (real) left-handed $Z$ coupling,
the $Z$ exchange contribution to $B_s$ mixing, which is negligible in the SM, cannot lead to sizable deviations from the SM. Concretely, in the considered scenarios we find, at 95\% C.L.,
\begin{align}
\text{left-handed mod.\ $Z$ couplings:}\qquad |S_{\psi\phi}-S_{\psi\phi}^\text{SM}|
& <0.008
\,,\\
\text{right-handed mod.\ $Z$ couplings:}\qquad |S_{\psi\phi}-S_{\psi\phi}^\text{SM}|
& <0.014
\,,\\
\text{generic mod.\ $Z$ couplings:}\qquad |S_{\psi\phi}-S_{\psi\phi}^\text{SM}|
& <0.04
\,.
\end{align}
Such NP contributions are well within the range allowed by the measurement of the $B_s$ mixing phase at LHCb~\cite{LHCb-CONF-2011-056}.

\subsubsection{Predictions for $b\to s\nu\bar\nu$ decays}

The two exclusive $b\to s\nu\bar\nu$ decays, $B \to (K,K^*) \nu\bar\nu$, and the inclusive one $B \to X_s \nu\bar\nu$ give access to four observables sensitive to NP: the three branching ratios and the $K^*$ longitudinal polarisation fraction $F_L$ in $B \to K^* \nu\bar\nu$. However, the observables are not all independent since they depend on only two real combinations of the complex Wilson coefficients $C_L$ and $C_R$~\cite{Grossman:1995gt,Melikhov:1998ug,Altmannshofer:2009ma},
\begin{equation}
\label{eq:epsetadef}
\epsilon = \frac{\sqrt{ |C_L|^2 + |C_R|^2}}{|(C_L)^{\rm SM}|}~, \qquad
\eta = \frac{-{\rm Re}\left(C_L C_R^*\right)}{|C_L|^2 + |C_R|^2}~.
\end{equation}

\noindent
For the central values of the hadronic parameters, one obtains\footnote{%
A lower central value for ${\rm BR}(B \to K \nu\bar\nu)$ is obtained if the experimental value of ${\rm BR}(B\to K\ell^+\ell^-)$ is used, assuming the latter decay to be SM-like \cite{Bartsch:2009qp}. Here we allow both decays to deviate from the SM prediction. We treat the $B\to\tau(\to K\bar\nu)\nu$ contribution \cite{Kamenik:2009kc} as a background to be subtracted from the experimental result.}
\cite{Altmannshofer:2009ma}
\beqa
\label{eq:epseta-BKsnn}
{\rm BR}(B \to K^* \nu\bar\nu) &=& 6.8 \times 10^{-6} \, (1 + 1.31 \,\eta)\epsilon^2~, \\
\label{eq:epseta-BKnn}
{\rm BR}(B \to K \nu\bar\nu)   &=& 4.5 \times 10^{-6} \, (1 - 2\,\eta)\epsilon^2~, \\
\label{eq:epseta-BXsnn}
{\rm BR}(B \to X_s \nu\bar\nu) &=& 2.7 \times 10^{-5} \, (1 + 0.09 \,\eta)\epsilon^2~, \\
\label{eq:epseta-FL}
 \langle F_L \rangle(B \to K^* \nu\bar\nu)             &=& 0.54 \, \frac{(1 + 2 \,\eta)}{(1 + 1.31 \,\eta)}~,
\eeqa
where $\langle F_L \rangle$ refers to the ratio of the branching ratio into a longitudinal $K^*$ over the total branching ratio. It can be extracted from the angular distribution of the $K^*\to K\pi$ decay products.

In the scenario with left-handed modified $Z$ couplings only, one has $\eta=0$, $F_L$ is SM-like and all the branching ratios are merely scaled by a common factor. We obtain, at 95\% C.L.,
\begin{equation}
\epsilon^2
=
\frac{{\rm BR}(B \to K^* \nu\bar\nu)}{{\rm BR}(B \to K^* \nu\bar\nu)_\text{SM}}
=
\frac{{\rm BR}(B \to K \nu\bar\nu)}{{\rm BR}(B \to K \nu\bar\nu)_\text{SM}}
 \in [0.5,1.3]
\,.
\end{equation}

\noindent
In the scenario with right-handed modified $Z$ couplings and SM-like left-handed couplings, one finds an anticorrelation between the two experimentally most promising modes, $B \to K \nu\bar\nu$ and $B \to K^* \nu\bar\nu$ (see \cite{Buras:2010pz,Straub:2010ih}). At 95\% C.L., we find
\begin{equation}
\frac{{\rm BR}(B \to K^* \nu\bar\nu)}{{\rm BR}(B \to K^* \nu\bar\nu)_\text{SM}}
\in[0.6,1.0] \,,~
\frac{{\rm BR}(B \to K \nu\bar\nu)}{{\rm BR}(B \to K \nu\bar\nu)_\text{SM}}
\in[1.0,2.6] \,,~
\frac{\langle F_L \rangle}{\langle F_L \rangle_\text{SM}}
\in[0.4,1.0] \,,
\end{equation}
The preference for a suppression of $F_L$ and ${\rm BR}(B \to K^* \nu\bar\nu)$ but an enhancement of ${\rm BR}(B \to K \nu\bar\nu)$ is again due to the negative values preferred for $Z_R$ commented on in section~\ref{sec:RHZ}. The large enhancement possible for ${\rm BR}(B \to K \nu\bar\nu)$  is close to the current experimental bound BR$(B^+ \to K^+ \nu\bar\nu) < 13 \times 10^{-6}$~\cite{delAmoSanchez:2010bk}.

\begin{figure}[tbp]
\centering
\includegraphics[width=0.417\textwidth]{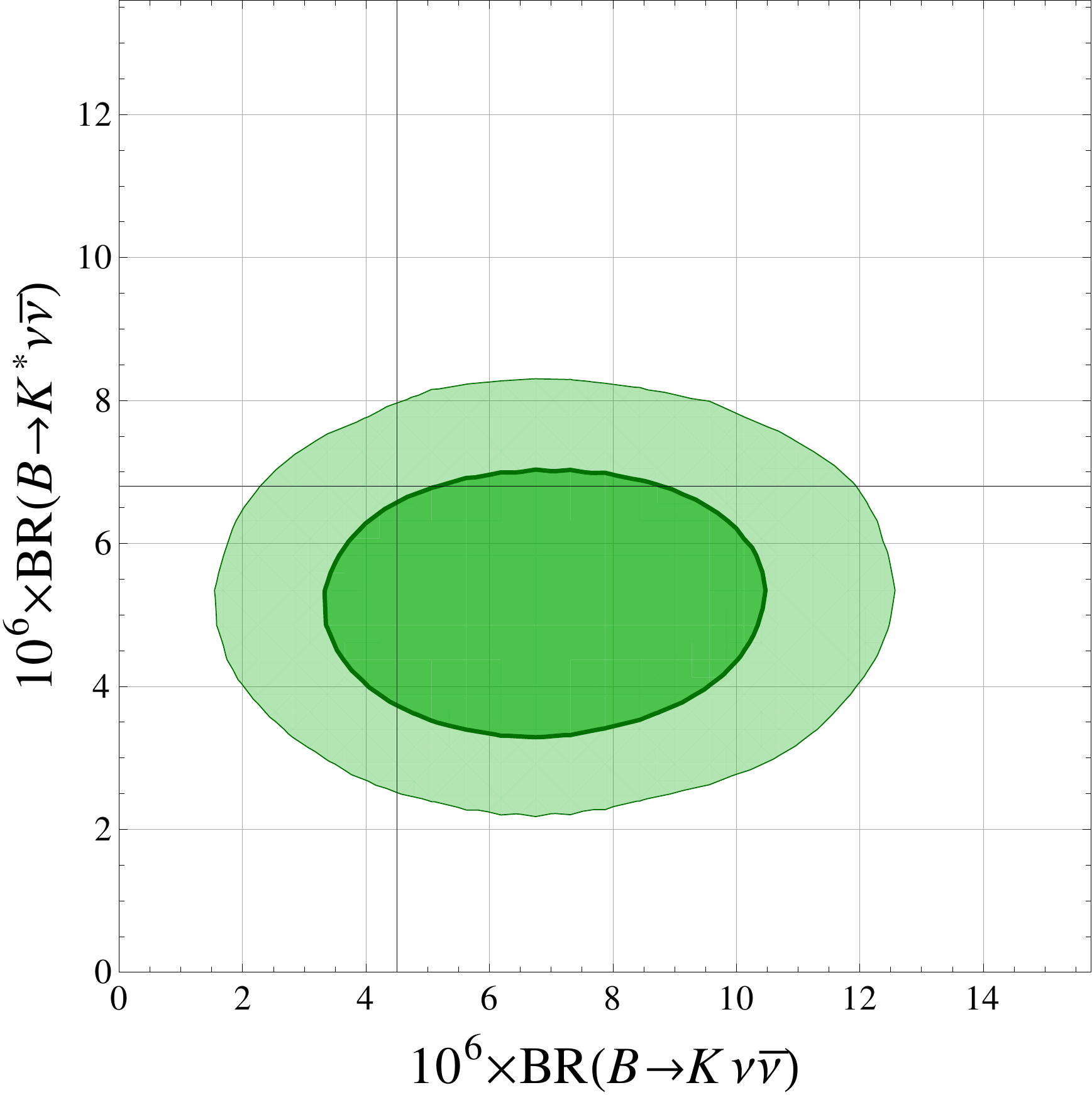}
\caption{Fit prediction for the branching ratios of $B\to K^{(*)}\nu\bar\nu$ for generic modified $Z$ couplings. Shown are 68\% and 95\% C.L. regions.}
\label{fig:bsnn}
\end{figure}

In the case of left- {\em and} right-handed $Z$ couplings, the correlation between the decays carries information on the size of left- vs. right-handed currents and is thus a valuable probe of the chirality structure of NP.  In figure~\ref{fig:bsnn}, we show the fit prediction for $B \to K \nu\bar\nu$ and $B \to K^* \nu\bar\nu$. We observe that the SM point is allowed at about 68\% C.L. Also in this case, only a small enhancement or a sizable suppression is allowed for the $B \to K^* \nu\bar\nu$ branching ratio, while an enhancement of the $B \to K \nu\bar\nu$ branching ratio up to a factor of 3 is possible.
For $F_L$, we find at 95\% C.L.
\begin{equation}
\frac{\langle F_L \rangle}{\langle F_L \rangle_\text{SM}}
\in[0.4,1.1] \,.
\end{equation}
In all the scenarios, the full allowed range of branching ratios can be probed at the next-generation $B$ factories~\cite{Meadows:2011bk}. 

Finally, we stress that these conclusions are only valid under the assumptions that modified $Z$ couplings dominate the NP contributions to $b\to s \nu\bar\nu$ and $b\to s\ell^+\ell^-$  semi-leptonic operators. Much larger effects in $B \to K^{(*)} \nu\bar\nu$ are possible in models where this is not the case, e.g. models with a non-universal $Z'$ coupling stronger to neutrinos than to charged leptons~\cite{Altmannshofer:2009ma}.\section{Conclusions}\label{sec:concl}

Rare decays with a $b\to s$ transition offer excellent opportunities to probe the flavour sectors
of extensions of the SM. The effects of new heavy degrees of freedom in these processes can be parametrised by modifications of the Wilson coefficients of local, non-renormalizable operators, which allows to constrain such NP effects in a model-independent way. In this work, we analysed the constraints on the Wilson coefficients that follow from the currently available experimental
data on $b\to s$ rare decays.
We took into account the measurements from the $B$ factories of the branching ratio of the radiative $B \to X_s \gamma$ decay, of the time-dependent CP asymmetry in $B \to K^* \gamma$, of the branching ratio of the inclusive $B \to X_s \ell^+ \ell^-$ decay, Belle and CDF data on the branching ratio and angular distribution of the exclusive $B \to K^* \mu^+ \mu^-$ decay and in particular the recent LHCb results on the branching ratio and angular distribution of $B \to K^* \mu^+ \mu^-$.

The constraints on the Wilson coefficients are obtained by using a $\chi^2$ function,
which depends on the Wilson coefficients and contains the theory predictions for the
observables and experimental averages as well as the corresponding uncertainties.

We have analysed the following scenarios where:
\begin{enumerate}
\item the dominant NP effects are captured already by one complex Wilson coefficient 
      or by a pair of real Wilson coefficients. This is a
      representative case of many NP models like the MSSM with MFV and flavour
      blind phases \cite{Altmannshofer:2008hc,Barbieri:2011vn}, non-MFV SUSY
      models~\cite{Altmannshofer:2009ne,Barbieri:2011ci,Barbieri:2011fc}
      and also models with dominance of $Z$ penguins;
\item the NP effects are accounted for by means of the full set of
      the 6 complex Wilson coefficients $C_{7,9,10}^{(\prime)}$.
\item the dominant NP effects in the semi-leptonic operators arise from non-standard 
      flavour changing $Z$ couplings.
\end{enumerate}

\noindent
While we refer to sections~\ref{sec:modelind} and~\ref{sec:zpeng} for a detailed description of
all our results, we want to emphasise here the following main messages:
\begin{itemize}
\item At the 95\% C.L., all best fit regions are compatible with the SM.
\item The combination of inclusive and exclusive $b\to s\ell^+\ell^-$
      observables exclude sign flips in various low-energy WCs.
      That is, the SM is likely to provide the dominant effects
      in low energy observables.
      In particular, we show that
\begin{itemize}
\item sign flips in $C_7$, $C_9$ {\em or} $C_{10}$ are excluded if NP enters dominantly through $Z$ penguins,
\item only a simultaneous sign flip of $C_7$, $C_9$ {\em and} $C_{10}$, which cannot be excluded by low-energy data alone, is allowed in the absence of non-standard CP violation or right-handed currents.
\end{itemize}
\item The new $A_{\rm FB}$ measurement at low $q^2$ from LHCb is already
      quite effective in constraining NP effects. The same is true for
      the time-dependent CP asymmetry in $B\to K^*\gamma$.
\item High-$q^2$ data on $B\to K^*\mu^+\mu^-$ are competitive with and complementary to the
      low-$q^2$ ones.
\end{itemize}

\noindent
Moreover, we have investigated the implications of the above constraints and the 
future prospects for observables in $b \to s \ell^+ \ell^-$ and $b\to s\nu\bar\nu$ 
transitions in view of improved measurements. In particular, we find that
\begin{itemize}
\item in the presence of non-standard CP violation, the low-$q^2$ angular CP asymmetries $A_7$ and $A_8$ in $B\to K^*\mu^+\mu^-$ can reach up to $\pm35\%$ and $\pm20\%$, respectively.
\item in the presence of right-handed currents, the low-$q^2$ angular observables $A_9$ and $S_3$ in $B\to K^*\mu^+\mu^-$ can reach up to $\pm15\%$.
\item in the absence of (pseudo)scalar currents, BR($\bsmm$) and BR($\bstt$) can be enhanced at most
by 50\% over their SM values, mainly due to the new measurement of $B\to K^*\mu^+\mu^-$
angular observables. In contrast, if scalar currents are at work, BR($\bsmm$) can still
saturate the current experimental bound and BR($\bstt$) can be enhanced by a factor of 3.
\item if NP in $b\to s\ell^+\ell^-$ and $b\to  s\nu\bar\nu$ processes is dominated by left-handed $Z$ penguins, an enhancement of the branching ratios of $B\to K^{(*)}\nu\bar\nu$ by more than 30\% is unlikely. If right-handed $Z$ penguins are present, $B\to K\nu\bar\nu$ can saturate the present experimental bound, while $B\to K^*\nu\bar\nu$ is unlikely to be enhanced.
\end{itemize}
The first two points highlight the importance of measuring observables in the  $B\to K^*\mu^+\mu^-$ angular distribution sensitive to right-handed currents and CP violation. Such measurements would be crucial to lift degeneracies in the space of Wilson coefficients which make it difficult at present to put strong constraints on individual coefficients in a completely generic NP model, as our analysis in sections \ref{sec:cMFV} and \ref{sec:genNP} showed.

In conclusion, the present work updates and generalises previous studies providing,
at the same time, a useful tool to test the flavour structure of any theory beyond
the SM.

\section*{Acknowledgements}

We thank Andrzej Buras and Thorsten Feldmann for helpful discussions and Christoph Bobeth and Zoltan Ligeti for useful comments.
Fermilab is operated by Fermi Research Alliance, LLC under Contract No. De-AC02-07CH11359 with
the United States Department of Energy. D.M.S. is supported by the EU ITN ``Unification in the
LHC Era'', contract PITN-GA-2009-237920 (UNILHC).

\appendix
\section{Statistical method}\label{sec:stat}

Here we give some details on our statistical method used to obtain the constraints on the Wilson coefficients in section~\ref{sec:globalfit} and \ref{sec:zpeng}. We use a Bayesian approach with the likelihood function
\begin{equation}
L(\vec C) = e^{-\chi^2(\vec C)/2} \,,
\end{equation}
where $\vec C$ is a 12-dimensional vector containing the real and imaginary parts of the 6 Wilson coefficients $C_{7,9,10}^{(\prime)}$ and the $\chi^2$ function has been defined at the beginning of section~\ref{sec:modelind}. We sample the posterior probability distribution, defined according to Bayes' theorem,
\begin{equation}
P(\vec C) = \frac{L(\vec C)\, \pi(\vec C)}{\int L(\vec C') \,\pi(\vec C')\,d\vec C'} \,,
\end{equation}
by means of a Markov Chain Monte Carlo (MCMC) analysis using the Metropolis-Hastings algorithm (see e.g. the review on statistics in \cite{Nakamura:2010zzi}). As proposal density, we use a multivariate Gaussian, whose width is optimised to tune the acceptance rate. We use a flat prior, $ \pi(\vec C)=1$, in the general case and a multivariate $\delta$ function in the restricted scenarios. The stationary density distribution of points in the Markov chain is proportional to $P(\vec C)$. Constraints on two-dimensional subspaces are obtained by projecting the points onto this plane. Predictions for observables presented in sections~\ref{sec:globalfit} and \ref{sec:zpeng} are obtained by evaluating the observable for central values of the theoretical input parameters at each point in the chain and interpreting their density distribution as posterior probability for the observable.

Two-dimensional confidence regions in sections~\ref{sec:globalfit} and \ref{sec:zpeng} are obtained by determining contours of constant posterior probability which contain 68\% (or 95\%) of the Markov chain points. Analogously, one-dimensional confidence regions are highest posterior density intervals, i.e. the posterior probability is higher everywhere inside the interval than outside, and they contain 68\% (or 95\%) of the points.

\section{Effective Wilson coefficients}\label{sec:ceff}

In sections~\ref{sec:modelind} and \ref{sec:zpeng}, we have put constraints on NP contributions to the Wilson coefficients at a matching scale of 160~GeV. Here we give the relation to the effective Wilson coefficients at low energies, which are the quantities relevant for the evaluation of observables.

In low-energy observables, the coefficients $C_7$ and $C_9$ always appear in a particular combination with four-quark operators (which can be found e.g. in \cite{Bobeth:1999mk}) in matrix elements. It hence proves convenient to define effective coefficients
$C_{7,9}^{\rm eff}$, which are given by \cite{Buras:1993xp}
\begin{align}
C_7^{\rm eff} & = C_7 -\frac{1}{3}\, C_3 -
\frac{4}{9}\, C_4 - \frac{20}{3}\, C_5\, -\frac{80}{9}\,C_6\,,
\label{eq:c7eff}\\
C_9^{\rm eff} & = C_9 + Y(q^2) \,,
\label{eq:c9eff}
\end{align} 
with
\begin{multline}
Y(q^2)  =  h(q^2,m_c) \left( \frac{4}{3}\, C_1 + C_2 + 6 C_3 + 60 C_5\right)
\\
-\frac{1}{2}\,h(q^2,m_b) \left( 7 C_3 + \frac{4}{3}\,C_4 + 76 C_5
  + \frac{64}{3}\, C_6\right)
\\
-\frac{1}{2}\,h(q^2,0) \left( C_3 + \frac{4}{3}\,C_4 + 16 C_5
  + \frac{64}{3}\, C_6\right)
+ \frac{4}{3}\, C_3 + \frac{64}{9}\, C_5 + \frac{64}{27}\,C_6 
\end{multline}
at leading order in $\alpha_s$ and $\Lambda_\text{QCD}/m_b$. Beyond the leading order, there are perturbative corrections as well as power corrections, which differ for inclusive and exclusive $b\to s\ell^+\ell^-$ decays.
We refer the reader to Refs. \cite{Beneke:2001at,Asatrian:2001de,Asatryan:2001zw,Ghinculov:2003qd,Huber:2005ig,Seidel:2004jh,Ligeti:2007sn,Huber:2007vv,Greub:2008cy} for these corrections. Ref. \cite{Beneke:2004dp} contains the expressions for the doubly Cabibbo-suppressed contribution to (\ref{eq:c7eff})--(\ref{eq:c9eff}) which is relevant for the SM prediction of the $B\to K^*\mu^+\mu^-$ CP asymmetries.

In the SM, one finds at the scale $\mu_b=4.8$\,GeV, to NNLL accuracy \cite{Altmannshofer:2008dz},
\begin{align}
C_7^{\rm eff}(\mu_b) & = -0.304
\,,&
C_9(\mu_b) & = 4.211
\,,&
C_{10}(\mu_b) & = -4.103
\,,
\end{align}
while the primed coefficients are negligible. Beyond the SM, but assuming the four-quark operators to be free of NP, one has
\begin{align}
C_7^{\rm eff}(\mu_b) & = C_7^{\rm eff, SM}(\mu_b) + C_7^\text{NP}(\mu_b)
\,,\\
C_9^{\rm eff}(\mu_b) & = C_9^{\rm eff, SM}(\mu_b) + C_9^\text{NP}
\,,\\
C_{10} & = C_{10}^{\rm SM} + C_{10}^\text{NP}
\,,\\
C_7'(\mu_b) & =C_7^{\prime\text{NP}}(\mu_b)
\,,\\
C_{9,10}'(\mu_b) & =C_{9,10}^{\prime\text{NP}}
\,.
\end{align}
While the NP contributions to $C_9^{(\prime)}$ and $C_{10}^{(\prime)}$ do not run, $C_7^{(\prime)}$ do and they mix with $C_8^{(\prime)}$ under renormalization. With leading order running, as is appropriate for NP contributions evaluated at one loop, from a high matching scale $\mu_h=160$~GeV, one finds
\begin{equation}
C_7^{(')\text{NP}}(\mu_b) = 0.623 ~C_7^{(')\text{NP}}(\mu_h) + 0.101 ~C_8^{(')\text{NP}}(\mu_h) \,.
\end{equation}
Since the low-energy observables are sensitive to $C_7^{(')\text{NP}}(\mu_b)$, the constraints we presented in sections~\ref{sec:modelind} and \ref{sec:zpeng} on $C_7^{(')\text{NP}}(\mu_h)$ for vanishing $C_8^{(')\text{NP}}$ can be interpreted as constraints on $C_7^{(')\text{NP}}(\mu_h) + 0.162 ~C_8^{(')\text{NP}}(\mu_h)$ for non-standard $C_8^{(\prime)}$.


\bibliographystyle{utphys}
\bibliography{bsll}

\end{document}